\documentclass[%
preprint,
superscriptaddress,
showpacs,
amsmath,amssymb,
aip,jcp,
floatfix,
citeautoscript,
]{revtex4-2}

\usepackage[export]{adjustbox}
\usepackage[caption=false]{subfig}
\usepackage{graphicx}
\usepackage{xspace}
\usepackage{textcomp}
\usepackage{siunitx}

\setcitestyle{super}

\usepackage{color}

\definecolor{grey}{rgb}{.65,.65,.65}

\begin{document}

\newcommand{\deu}{$^{2}$H\xspace}
\newcommand{\hyd}{$^{1}$H\xspace}
\newcommand{\Al}{$^{27}$Al\xspace}
\newcommand{\hw}{D$_2$O\xspace}
\newcommand{\w}{H$_2$O\xspace}
\newcommand{\VV}{$V_{zz}$\xspace}
\newcommand{\cq}{$C_\mathrm{Q}$\xspace}

\newcommand{\texp}[1]{\ensuremath{\times 10^{#1}}}


\title{Orientation and dynamics of water molecules in beryl}

\author{Vojt\v{e}ch Chlan}
\email{vojtech.chlan@mff.cuni.cz}
\affiliation{Charles University, Faculty of Mathematics and Physics, Department of Low Temperature Physics, V Hole\v{s}ovi\v{c}k\'ach 2, 180 00 Prague 8, Czech Republic}
\author{Martin Adamec}
\affiliation{Charles University, Faculty of Mathematics and Physics, Department of Low Temperature Physics, V Hole\v{s}ovi\v{c}k\'ach 2, 180 00 Prague 8, Czech Republic}
\affiliation{Institute of Physics of the Czech Academy of Sciences,	Na Slovance 2, 182 00 Prague 8, Czech Republic}
\author{Helena \v{S}t\v{e}p\'{a}nkov\'{a}}
\affiliation{Charles University, Faculty of Mathematics and Physics, Department of Low Temperature Physics, V Hole\v{s}ovi\v{c}k\'ach 2, 180 00 Prague 8, Czech Republic}
\author{Victor G.\ Thomas}
\affiliation{V.S. Sobolev Institute of Geology and Mineralogy SB RAS, 630090 Novosibirsk, Russia}
\author{Filip Kadlec}
\affiliation{Institute of Physics of the Czech Academy of Sciences,
Na Slovance 2, 182 00 Prague 8, Czech Republic}

\begin{abstract}
Behavior of individual molecules of normal and heavy water in beryl single crystals was studied by \hyd and \deu nuclear magnetic resonance spectroscopy. From temperature dependences of the spectra we deduce that type-I water molecules
embedded in the beryl voids are oriented quite differently from the view established in the literature. Namely, contrary to earlier assumptions, their H-H lines deviate by about \qty{18}{\degree} from the hexagonal axis. We suggest that this is due to the molecules attaching to the oxygen atoms forming the beryl structural voids by a hydrogen bond. Our analysis shows that the molecules perform two types of movement: (i) rapid librations around the axis of the
hydrogen bond, and (ii) less frequent orientational jumps among the twelve possible binding sites in the beryl voids. The frequencies of the librational motions are evaluated from a simple thermodynamic model, providing a good quantitative agreement with the frequencies of librations from optical experiments reported earlier.
\end{abstract}

\maketitle
\section{Introduction}

\subsection{Water confined in beryl} 
\begin{figure}
    \centering
    \includegraphics[width=0.95\columnwidth]{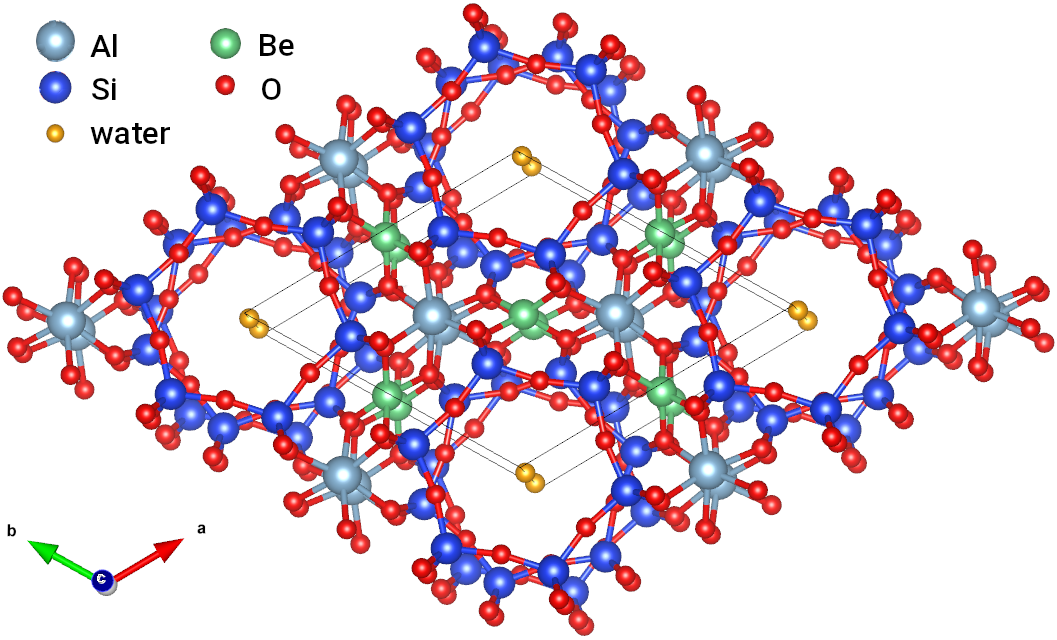}
    \caption{View of the crystal structure of beryl along the hexagonal axis
    $c$. The thin lines denote the unit cell whereas the yellow circles indicate
  the possible positions of the water molecules within the structural voids. The structural data\cite{Gibbs68} was visualized using the Vesta software.\cite{Momma11} } 
    \label{fig:struct}
  \end{figure}

Water confined in nanoscale volumes
manifests many unusual properties and has recently drawn a considerable
attention. Among the known corresponding structures,
hydrated beryl Be$_3$Al$_2$Si$_6$O$_{18}$ is a system attracting a broad interest where tendencies to low-temperature ferroelectric ordering of the crystal water were clearly demonstrated \cite{Gorshunov14,Gorshunov16}. In fact, the water molecules can occupy regularly spaced crystal sites which are  enclosed by oxygen atoms (see Fig.~\ref{fig:struct}). These sites define well  the molecules' positions; in contrast, their angular orientations are  generally variable. Note also that, as a rule, only a partial beryl hydration is  achieved. Thus, in this system, unlike in common condensed phases of water,  hydrogen bonding among the water molecules is suppressed, and they interact  predominantly via electric dipole--dipole interactions involving their dipole moments  of \qty{1.85}{D} (\qty{6.17e-30}{\coulomb\metre}). Incipient ferroelectricity has  been documented especially by observations of collective vibrations of the  water molecules, producing a ferroelectric soft phonon mode. This soft mode was  observed in the THz-range spectra of dielectric permittivity, obeying the  usual Curie-Weiss and Cochran temperature dependences \cite{Gorshunov16}. In  the reported case, a negative Curie temperature of $T_{\rm C}\!\approx\,$\qty{-20}{\kelvin}
  was determined, so no ferroelectric state could be achieved. Additionally, at
temperatures below about \qty{20}{\kelvin}, the phonon no more softened; instead, its frequency was leveling off, which has been attributed to quantum tunneling
\cite{Kolesnikov2016,Zhukova14}. For the above reasons, hydrated beryl has been
one of the most studied crystal systems featuring confined
water
\cite{Belyanchikov2017,Gorshunov14,Gorshunov16,Zhukova14,Finkelstein17,Arivazhagan17,Kolesnikov2016}.
Owing to its well defined geometry and interesting observed phenomena,
hydrated beryl can serve as a model structure for more detailed spectroscopic
and theoretical studies, with the aim of improving the current
  knowledge of the underlying phenomena which may favor or suppress the ordering
  of confined water molecules.

The crystal structure of beryl is hexagonal (space group $P6/mcc$), and it
contains channels of voids running along its hexagonal axis.  Each of
these voids may accommodate one water molecule.
For more than fifty years, the water molecules have been supposed to take up two possible types of orientations within the voids,
as hypothesized first by Wood and Nassau based on their infrared spectra
  analysis. \cite{Wood67} They concluded that the molecules will orient
themselves with the H--H lines oriented either parallel to
the hexagonal axis (``type-I water''), or perpendicular to it (``type-II
water''); the latter type should be present especially in crystals with an
additional doping, as the oxygen may bind to an impurity atom located within
the channel. The earlier studies dealing with the interactions among the water
molecules and their collective dynamics assumed the type-I molecules to rotate
around the hexagonal axis of beryl, so the molecules' planes remained parallel
to the hexagonal axis. At the same time, it was supposed that the molecules are
subjected, in their angular orientations, to a local potential exhibiting six
equivalent minima separated by angles of \qty{60}{\degree}.
\cite{Belyanchikov2017,Kolesnikov2016,Dressel18,Finkelstein17,Zhukova14} Whereas the properties of the molecular
ensemble were studied quite extensively---see the above
references---, the orientations and dynamics
of the individual molecules in the voids have been still less
explored, despite being crucial for the bulk properties. Such local,
molecular-level information can be provided namely by Nuclear Magnetic
Resonance (NMR) spectroscopy.

\subsection{NMR spectroscopy of water in beryl}
Water molecules are expected to produce a single
peak in the \hyd NMR spectrum, because the spin of the \hyd nucleus (proton) is \textonehalf\ and both hydrogen atoms are equivalent due to symmetry. The peak may become split and/or shifted when anisotropic interactions are involved, e.g., dipolar interactions or anisotropy of chemical shielding. In cases when the water molecules are highly mobile, such as in liquid or gas phases, the effect of these anisotropic interactions on the NMR spectrum gets averaged by the fast molecular reorientations (\textit{motional narrowing}). Then, the \hyd nuclei in water molecules are exposed to the same, averaged local fields and the \hyd NMR spectrum of water consists of a single, usually very narrow peak.\cite{Abragam1961}

Water molecules in solids have their dynamics significantly restricted. When the molecules are static or their reorientations are very slow (e.g., in ice or as a water of crystallization), the anisotropic interactions are not fully averaged and the NMR spectrum comprises contributions from all arrangements of the molecule present in the sample. For single crystals, the NMR spectra depend on the crystal orientation with respect to $\mathbf{B}_0$, while the NMR spectra of powders with randomly oriented grains display powder-pattern features.\cite{Abragam1961} In terms of magnitude, the anisotropy of chemical shielding 
of \hyd in water molecule ranges from 19 to \qty{35}{ppm} for various phases of water\cite{Modig2002}, which for $B_0\sim\,$\qty{10}{\tesla} corresponds to shifts or splittings of peaks in the \hyd spectrum of about \qty{10}{\kHz}. The effect of dipolar interactions between \hyd nuclei within the water molecule may be an order of magnitude larger, the dipolar interaction is thus often the dominant source of anisotropy in \hyd NMR spectra of solids.\cite{Abragam1961}

The character of the \hyd NMR spectrum of water molecules enclosed in the crystal voids is different from the two cases presented above. The confined molecules are not static as in ice, and the molecular reorientations are not isotropic as in liquid water. Thus, the anisotropic interactions are not fully averaged and the \hyd NMR spectrum of water consists of more than one peak.
This is the case of water molecules confined in the voids of beryl
crystal, which was first studied using NMR by Paré and Ducros\cite{Pare1964} and
by Sugitani \textit{et al.}\cite{Sugitani1966} already in 1960s. In both these works, the \hyd NMR spectra comprise doublet of peaks arising due to the nuclear dipolar interaction between the \hyd nuclei within the water molecules. The signal was attributed to type-I water. Moreover, the observed dipolar splitting increased linearly with decreasing temperature and saturated below 100\,K. The natural interpretation then was that the water molecules oscillate around their equilibrium positions when H--H line is parallel to the hexagonal axis, and that the amplitudes of oscillations increase with temperature\cite{Pare1964}.

The idea that the type-I water molecules are simply oscillating around the
hexagonal axis is, however, not entirely correct, since even at the lowest
temperatures the observed dipolar splitting does not reach the expected value
which is precisely given by the distance between the \hyd nuclei within the water
molecule. As we show in this paper, the movements of water molecules must be more
complex. This deduction is based on measuring and analyzing the dipolar
splitting in \hyd NMR quantitatively, and especially by measuring and analyzing  \deu NMR in beryl hydrated by heavy water.

The spin quantum number of the \deu nucleus equals 1 and so, in the presence of an
external magnetic field, two transitions between the nuclear energy levels of the Zeeman multiplet are observable. Thus, the \deu NMR spectrum of heavy water in beryl is expected to comprise two doublets of peaks, one for each \deu nucleus. In contrast to the \hyd case, the \deu dipolar interaction is small and a quadrupolar splitting occurs due to an interaction between the electric quadrupole moment of the \deu nucleus and the gradient of surrounding electric fields. The quadrupole splitting depends on the direction of the external magnetic field vector $\mathbf{B}_0$ with respect to the axes of the electric field gradient (EFG) tensor. Whereas in the \hyd case it is the orientation of the H--H line that determines the magnitude of dipolar splitting, in case of \deu the orientation of the O--H bond becomes important instead, since in the \w molecule the principal axis of the EFG tensor at the hydrogen site points approximately along the O--H bond. This is an essential difference between the NMR interactions in \w and \hw that
allows for extracting more complete information about the orientation of water
molecule in the beryl voids: the isotopes \hyd and \deu serve as two completely
different probes. Since the values of dipolar splitting in \hyd and
quadrupole splitting in the \deu NMR spectra depend significantly on the orientations of water molecules, we can evaluate the orientation and dynamics of the water molecules by analyzing the temperature dependences of the measured dipolar and quadrupolar splittings.

\section{Methods}

Four high-quality synthetic beryl single crystals were studied: two samples  containing normal water and two containing heavy water. The samples were grown from oxides on seed plates at \qty{600}{\degreeCelsius} and \qty{1.5}{kbar} in a hermetically sealed gold vessel by the hydrothermal method of Thomas and Klyakhin\cite{Thomas1987}. All samples were cut into cylinders about \qty{2}{\mm} in diameter, with one sample for each hydrogen isotope having the hexagonal axis oriented parallel to the axis of cylinder and the other perpendicular to it. All beryl samples contained type-I water predominately; because of the artificial origin of the samples, the content of alkali atoms and other such impurities and hence also the content of type-II water was negligible.

NMR spectra of water-containing beryl single crystals were acquired in a magnetic field of \qty{9.41}{\tesla} (Larmor frequency $f_0 =\,$\qty{400.185}{\MHz} for \hyd and \qty{61.431}{\MHz} for \deu) using a Bruker Avance II
spectrometer. Solid echo pulse sequences  $90^\circ_x\!-\!\tau\!-\!90^\circ_y\!-\!\tau$ (with
lengths of the $90^\circ$ pulses 0.5--\qty{1.2}{\us} for \hyd and
3.5--\qty{5.0}{\micro\second} for \deu, and with a delay of $\tau\!\approx\!60$--\qty{100}{\us}) were applied to excite the NMR signal in order to ensure proper refocusing of the spin magnetization in the presence of strong nuclear dipolar
interaction (\hyd case) or electric quadrupole interaction (\deu case). The
trigger delays between the subsequent scans were adjusted at each temperature point in order to compensate the strong temperature dependence of the spin-lattice relaxation time. Its value amounted to about \qty{1}{\second} and \qty{20}{\second} for \hyd and \deu, respectively, at the highest temperatures, and to more than \qty{100}{\second} for both isotopes at the lowest temperatures.
At each temperature point, 2--64 scans (\hyd spectra) or 8--1024 scans (\deu spectra) were performed, and the acquired spin echo signals were coherently summed.

The temperature dependences were measured in a Janis helium continuous flow cryostat, and the temperature was monitored with Cernox CX-1030 temperature sensor in the proximity of the NMR coil. The orientation of the beryl crystal with respect to the external magnetic field was adjusted carefully, with an estimated error of less than \qty{3}{\degree}. Using a goniometer, the angular dependence of \deu spectra was measured at room temperature.

Special care was taken to keep parasitic \hyd signals at the minimum level. We used a custom-made NMR probe without any plastic or Teflon parts, and an NMR coil made from unvarnished silver wire. The probe and the sample were dehumidified prior to the experiments. The remaining small parasitic \hyd signal near Larmor frequency most probably originated from the cap of the temperature sensor.

\section{Results}

\hyd and \deu NMR spectra were measured in the temperature range 5--\qty{360}{\kelvin}, and for each isotope two cylindrical samples were studied: in one, the hexagonal crystallographic axis was parallel to the cylinder axis and in the other perpendicular to it. We first describe and interpret the room temperature spectra of \hyd and \deu, then we deal with their temperature dependences in the range of approximately 5--\qty{360}{\kelvin}.	
\begin{figure}
	\includegraphics[width=0.5\columnwidth]{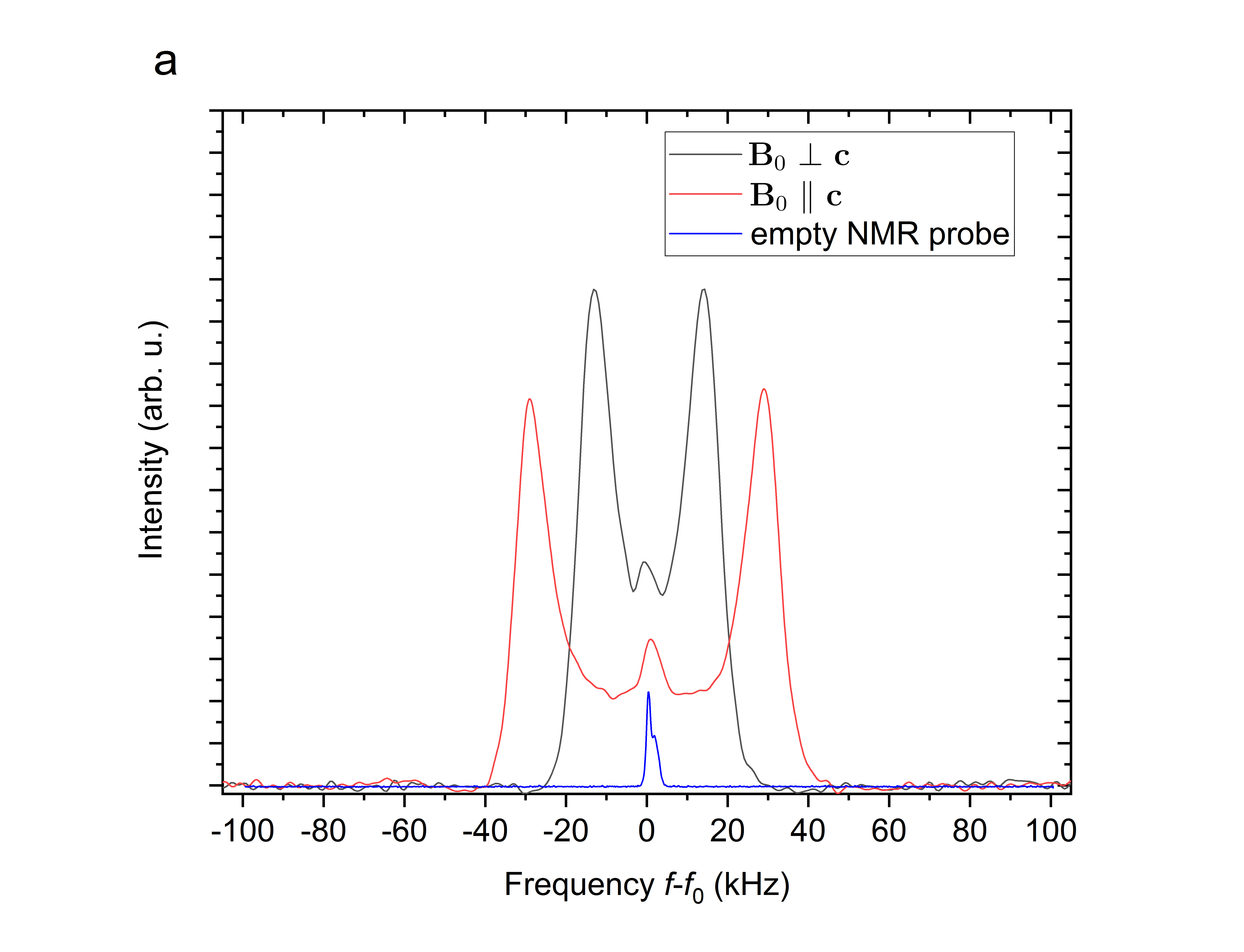}\includegraphics[width=0.5\columnwidth]{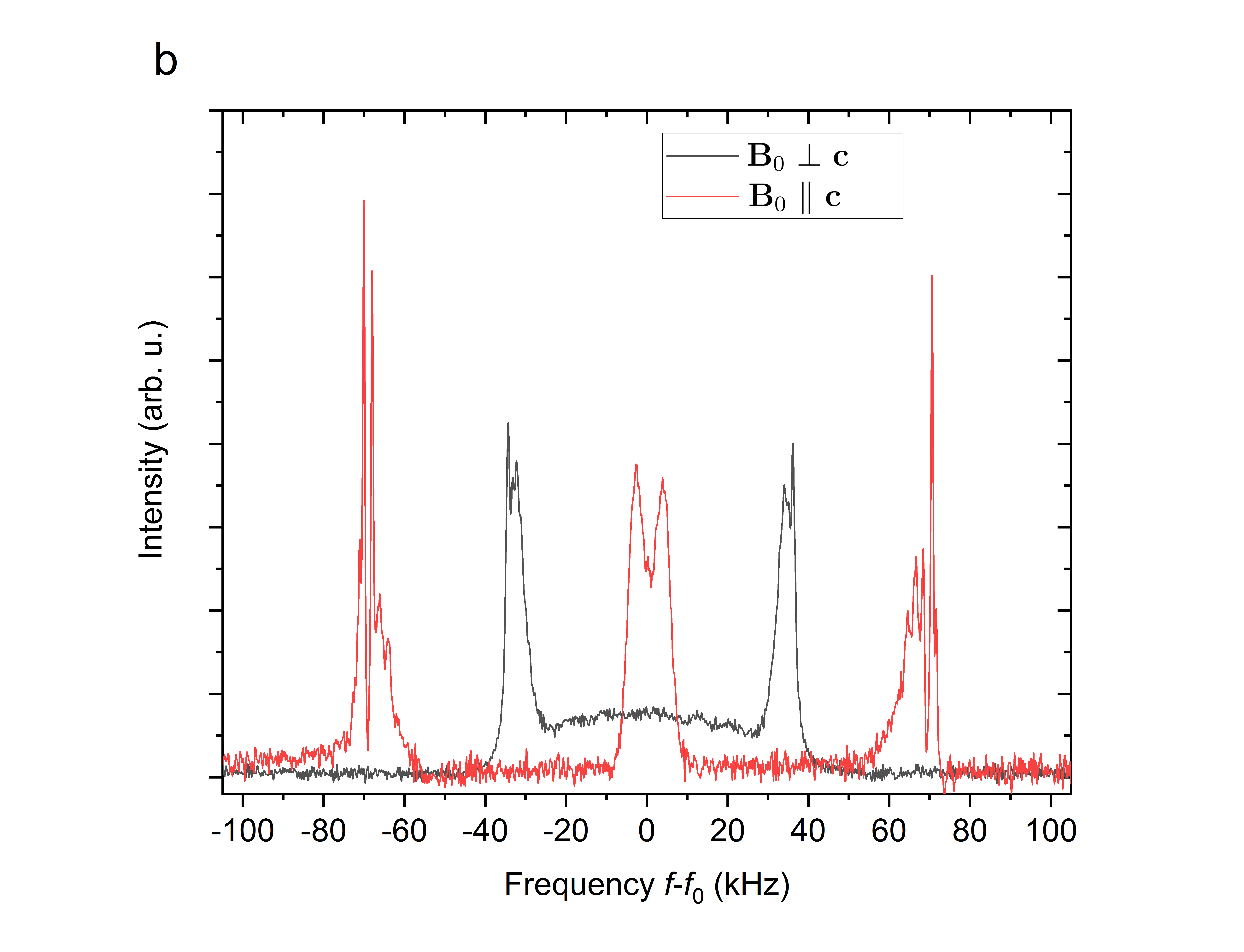}
	\caption{\label{fig:spectraRT} \hyd (a) and \deu (b) NMR spectra of water in beryl measured at room temperature under two experimental settings, with the external magnetic field $\mathbf{B}_0$ oriented perpendicular and parallel with respect to the hexagonal axis $\mathbf{c}$ of the crystal, respectively.  Near $f\!=\!f_0$, the \hyd spectra contain a small parasitic signal originating from the NMR probe.}
\end{figure}

\subsection{\hyd NMR spectra at room temperature}
The room temperature \hyd NMR spectrum of water molecules in beryl
(Fig.~\ref{fig:spectraRT}a) consists of a doublet of peaks, originating in the magnetic dipolar interaction between the two \hyd nuclei in the water molecule. The value of the dipolar splitting $\Delta_{\rm D}$ depends on the dipolar constant $\delta$ as well as on the orientation of the line connecting the interacting nuclei with respect to the direction of the external magnetic field $\mathbf{B}_0$, which is described by a polar angle $\vartheta_\mathrm{D}$\cite{Abragam1961}:
\begin{eqnarray}
  \Delta_\mathrm{D} = \delta \left( 3 \cos^2 \vartheta_\mathrm{D} - 1 \right), \,\,\,\, \delta=\frac{3}{2}\frac{\mu_0}{4\pi}\frac{\gamma^2\hbar}{r^3}\,.
\label{eq:dipolar}
\end{eqnarray}
The dipolar constant $\delta$ depends on the gyromagnetic ratio $\gamma$ of the interacting nuclei, and it decays with the third power of their mutual distance $r$. Clearly, the dipolar splitting $\Delta_\mathrm{D}$ in the measured spectra is different for the two displayed orientations of the crystal with respect to the external magnetic field, as the parallel orientation ($\mathbf{B}_0 \parallel \mathbf{c}$) corresponds to $\vartheta_\mathrm{D}\!\approx\!0^\circ$ and the perpendicular one ($\mathbf{B}_0 \perp \mathbf{c}$) to $\vartheta_\mathrm{D}\approx90^\circ$. The larger splitting observed for the parallel orientation implies that the lines connecting the hydrogen atoms (H--H) in the water molecules are preferentially oriented nearly along the hexagonal axis of beryl crystal, i.e., type-I water is observed in the \hyd NMR spectra. No signals corresponding to type-II water sites were detected, which is in accord with the supposed negligible concentration of alkali metal atoms in the studied synthetic beryl crystals.

The room temperature values of the observed dipolar splitting
$\Delta_\mathrm{D}\!\approx\,$\qty{59}{\kHz} and $\approx\,$\qty{29}{\kHz} for the parallel and
perpendicular orientations, respectively, are significantly smaller
than expected for a dipolar interaction between two \hyd nuclei in a
type-I oriented water molecules. In fact, the dipolar constant
$\delta$ between two \hyd nuclei is given [see Eq.~(\ref{eq:dipolar})] by their distance $r$,
which is relatively well known ($r\!\approx\,$\qty{1.524}{\angstrom}). An orientation with the
 H--H line parallel ($\vartheta_\mathrm{D}\!=\!0^\circ$) or
perpendicular ($\vartheta_\mathrm{D}\!=\!90^\circ$) to the external field would
therefore yield a splitting of the doublet  $\Delta_\mathrm{D}=\,$\qty{117}{\kHz}
and \qty{58.7}{\kHz}, respectively. Such discrepancy can be explained by (i) the actual
value of the angle $\vartheta_\mathrm{D}$ being appreciably different from 0
or 90$^\circ$, or (ii) by an averaging of the dipolar interaction strength due to rapid molecular motions. In the further text (section \ref{analysis}), we show that both mechanisms are responsible for the reduction of the observed dipolar splitting.

Additionally, the \hyd nuclei in the water molecules are affected by
dipolar interactions with other nuclear species in the surroundings, mainly
the \hyd nuclei of other water molecules in the neighboring
voids and \Al nuclei in the beryl crystal structure. However, these
distant partners induce much weaker dipolar fields, and thus these
interactions only lead to a broadening of the resonance peaks. Also, despite
our efforts to avoid any hydrogen-containing materials in the vicinity of
the radiofrequency coil, a small parasitic \hyd line was detected close to
the Larmor frequency. This originated probably from hydrogen atoms present in a small amount in the body of the NMR probe, or in the coating of the
Cernox temperature sensor.

\subsection{\deu NMR spectra at room temperature}
The \deu NMR spectrum of beryl containing deuterated water is more complex. When the hexagonal axis of beryl crystal is oriented parallel to the external magnetic field, a pair of well resolved doublets is observed in the \deu spectrum: an outer doublet with a larger splitting of $\approx\,$\qty{140}{\kHz} and an inner doublet with a splitting of $\approx\,$\qty{7}{\kHz}. In contrast to the \hyd case, these doublets are formed as a result of the electric quadrupole interaction at each \deu nucleus. The presence of two distinct quadrupole doublets in the spectrum implies that the two \deu nuclei in the water molecules are non-equivalent. This in turn means that the motion of water molecules cannot be described by simple oscillations about the hexagonal axis because, in such a case, the averaged local fields would be identical for both \deu nuclei. Similarly to the dipolar splitting in the \hyd case, the electric quadrupole splitting  $\Delta_\mathrm{Q}$ also depends on the orientation of the water molecules with respect to the external magnetic field, yet in a more complex way:\cite{Abragam1961,Freude2000}

\begin{eqnarray}
	\Delta_\mathrm{Q} &=&\frac{3}{4} C_\mathrm{Q} \left( 3 \cos^2 \vartheta_\mathrm{Q} - 1 + \eta \sin^2 \vartheta_\mathrm{Q} \cos 2\varphi_\mathrm{Q} \right), \\ C_\mathrm{Q} &=&\frac{eQV_{zz}}{h}.
	\label{eq:quad}
\end{eqnarray}
The polar $\vartheta_\mathrm{Q}$ and azimuthal $\varphi_\mathrm{Q}$ angles characterize the direction of the external magnetic field $\mathbf{B}_0$ within the principal axis system (PAS) of the EFG tensor $V$, which is described by two parameters:
the largest component \VV, $|V_{zz} |\geq|V_{yy} |\geq|V_{xx} |$, and a
dimensionless asymmetry factor $\eta=\frac{V_{xx}-V_{yy}}{V_{zz}}$, $0\leq\eta\leq1$.
For the hydrogen site in the water molecule, the principal
axis belonging to \VV of the tensor $V$ is pointing approximately along
the O--H bond, and $V_{yy}$ is perpendicular to the \w molecular plane (Fig.~\ref{fig:geom}a). The electric quadrupole interaction constant $C_\mathrm{Q}$ is given by the nuclear
quadrupole moment of \deu, $Q\!=\,$\qty{2.85783(30)}{mb} (milibarn, \qty{1}{mb}\,=\,\qty{e-31}{\metre\squared})
\cite{Pyykk2018}, and by the value of \VV at the \deu nucleus, which depends
significantly on the phase of water. In the gas phase, $C_\mathrm{Q}\!=\,$\qty{307.5(6)}{\kHz}
with $\eta\!=\!0.138(1)$\cite{Garvey1977,Bluyssen1967}, in liquid water
$C_\mathrm{Q}\!=\,$\qty{250(7)}{\kHz}\cite{Struis1987,Gordalla1986} ($\eta$ is not known in
liquid, but its value is estimated to be similar to those in other phases), and
$C_\mathrm{Q}\!=\,$\qty{213.4(3)}{\kHz} with $\eta\!=\!0.112(5)$ in ice
I$_\mathrm{h}$\cite{Edmonds1975}. The values of $C_\mathrm{Q}$ for \deu in
liquid and ice are reduced compared to the gas phase due to numerous interactions
with neighboring molecules via hydrogen bonds in the condensed phases. The
quadrupole constant $C_\mathrm{Q}$ of \deu in heavy water confined within the
beryl voids is unknown, but its value can be expected roughly around
200--\qty{300}{\kHz}. The magnitude of the observed quadrupole splitting at room
temperature (Fig.~\ref{fig:spectraRT}b) is again diminished due to molecular
motions, similarly as in the case of \hyd dipolar splitting in \w.

The dipolar interaction within the water molecule, which is the dominating
interaction in the \hyd spectra, is much weaker in case of \deu due to the
magnetic moment of the \deu nucleus being $6.5\times$ smaller than the
moment of \hyd. The intramolecular \deu--\deu dipolar interactions thus manifest
only as a fine splitting ($\approx\,$\qty{2}{\kHz}) of the spectral lines, which is well
noticeable for the peaks of the outer quadrupole doublet [see Fig.~\ref{fig:spectraRT}(b)], and any further dipolar interactions may be neglected.

\begin{figure}
	\includegraphics[width=0.5\columnwidth]{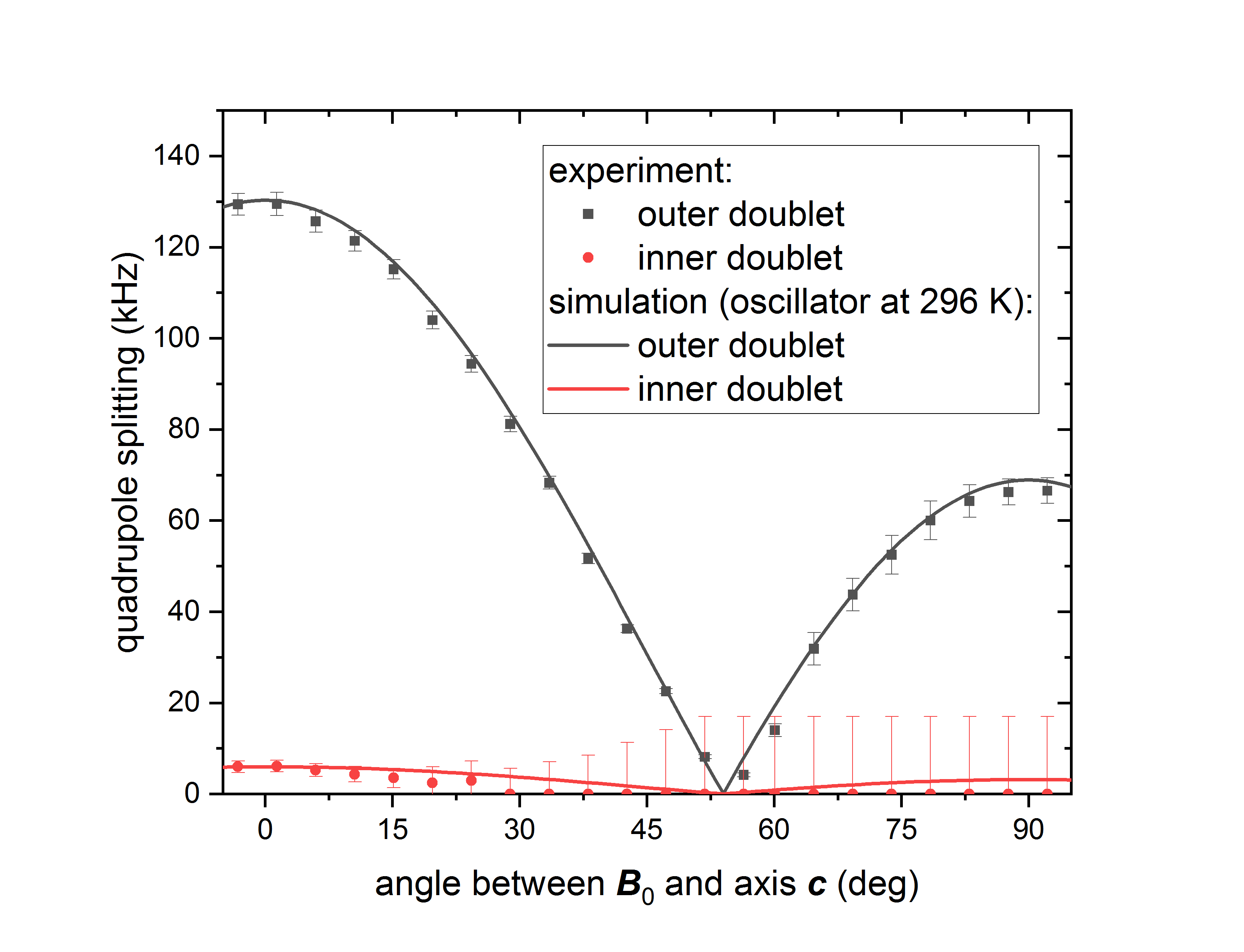}
	\caption{\label{fig:angular} Angular dependences of the quadrupole splittings $\Delta_{\rm Q}$ obtained from the \deu NMR spectra of a deuterated beryl single crystal at room temperature. Symbols denote experimental data points, solid lines represent dependence simulated using linear harmonic oscillator model described in section \ref{analysis}.}
\end{figure}

In the $\mathbf{B}_0 \perp \mathbf{c}$ orientation
(Fig.~\ref{fig:spectraRT}b), the \deu NMR spectrum also consists of two
doublets, however, only the outer one is well resolved. The peaks of the inner doublet with a smaller splitting overlap and they are severely broadened,
especially at lower temperatures. This behavior can be well observed in the angular dependence of the \deu quadrupole splitting $\Delta_{\mathrm Q}$ measured for several angles between the two extreme cases $\mathbf{B}_0 \perp \mathbf{c}$ and $\mathbf{B}_0\parallel \mathbf{c}$ (Fig.~\ref{fig:angular} right). The peaks of the inner doublet are resolved only for angles between $\mathbf{B}_0$ and $\mathbf{c}$ below $30^\circ$. The excessive broadening is caused by insufficient motional averaging with respect to the hexagonal axis, as we explained below in the section \ref{analysis}.

\subsection{Temperature dependence of \hyd and \deu NMR spectra} 
\begin{figure}
\includegraphics[width=0.5\columnwidth]{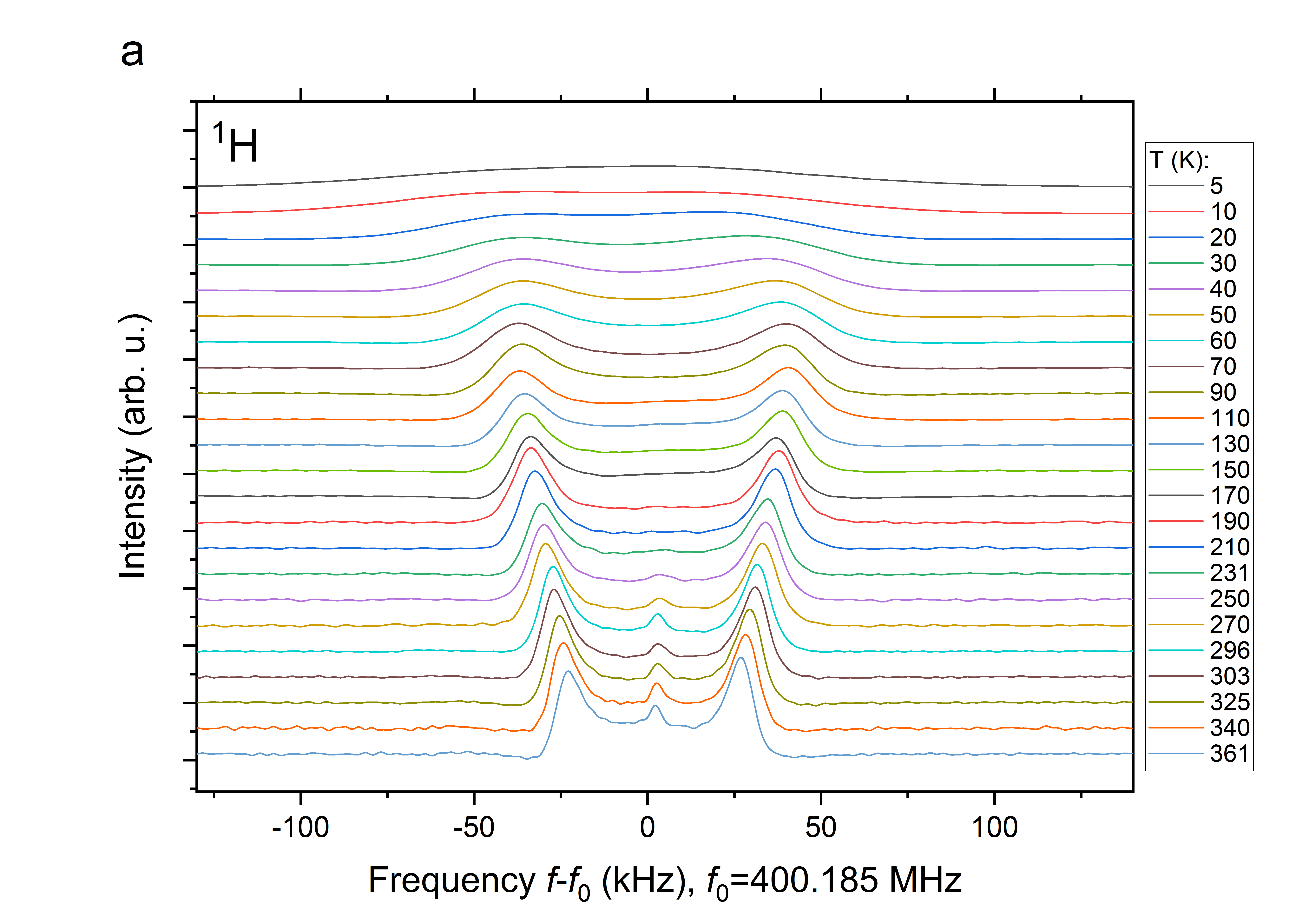}\includegraphics[width=0.5\columnwidth]{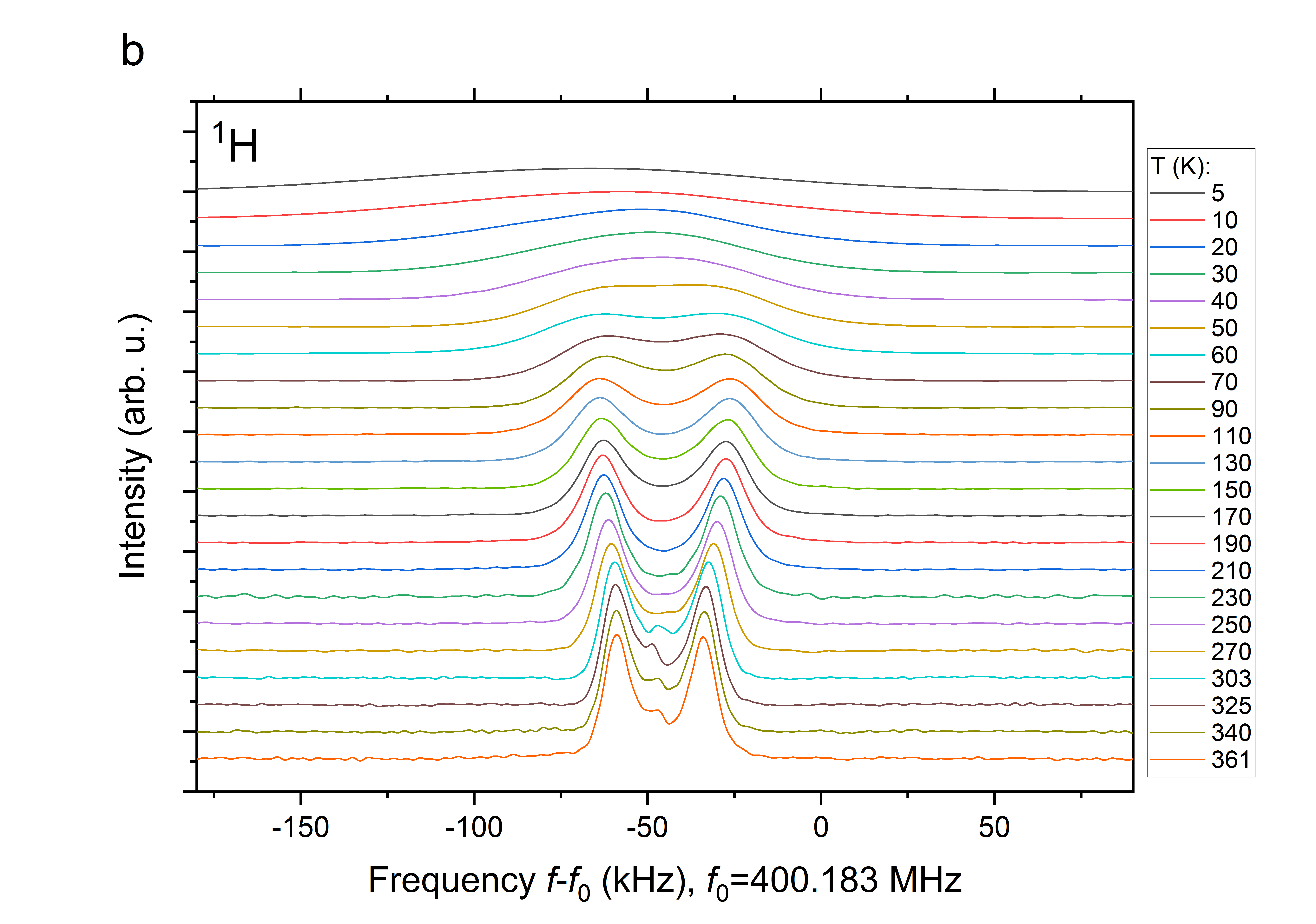}
\includegraphics[width=0.5\columnwidth]{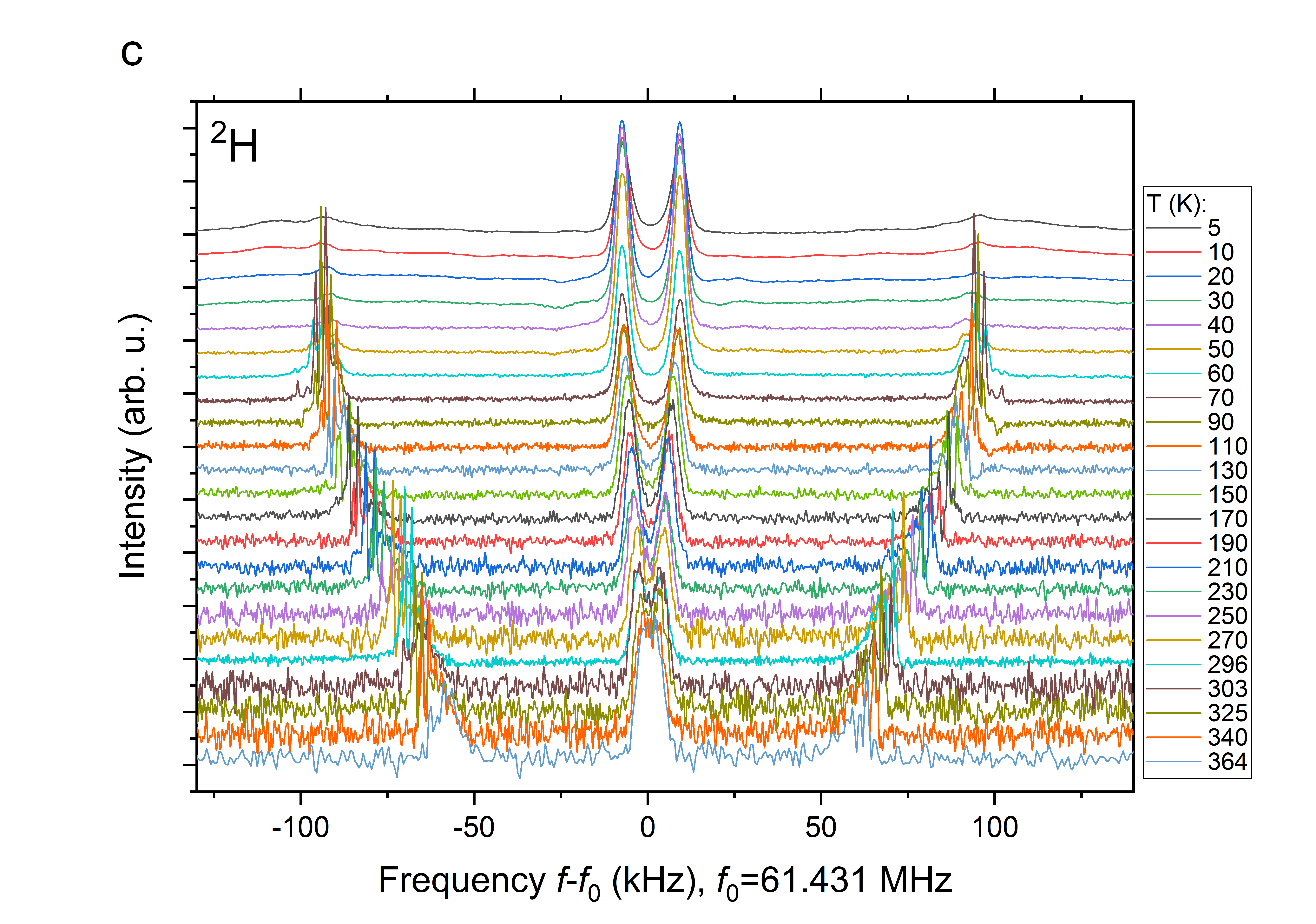}\includegraphics[width=0.5\columnwidth]{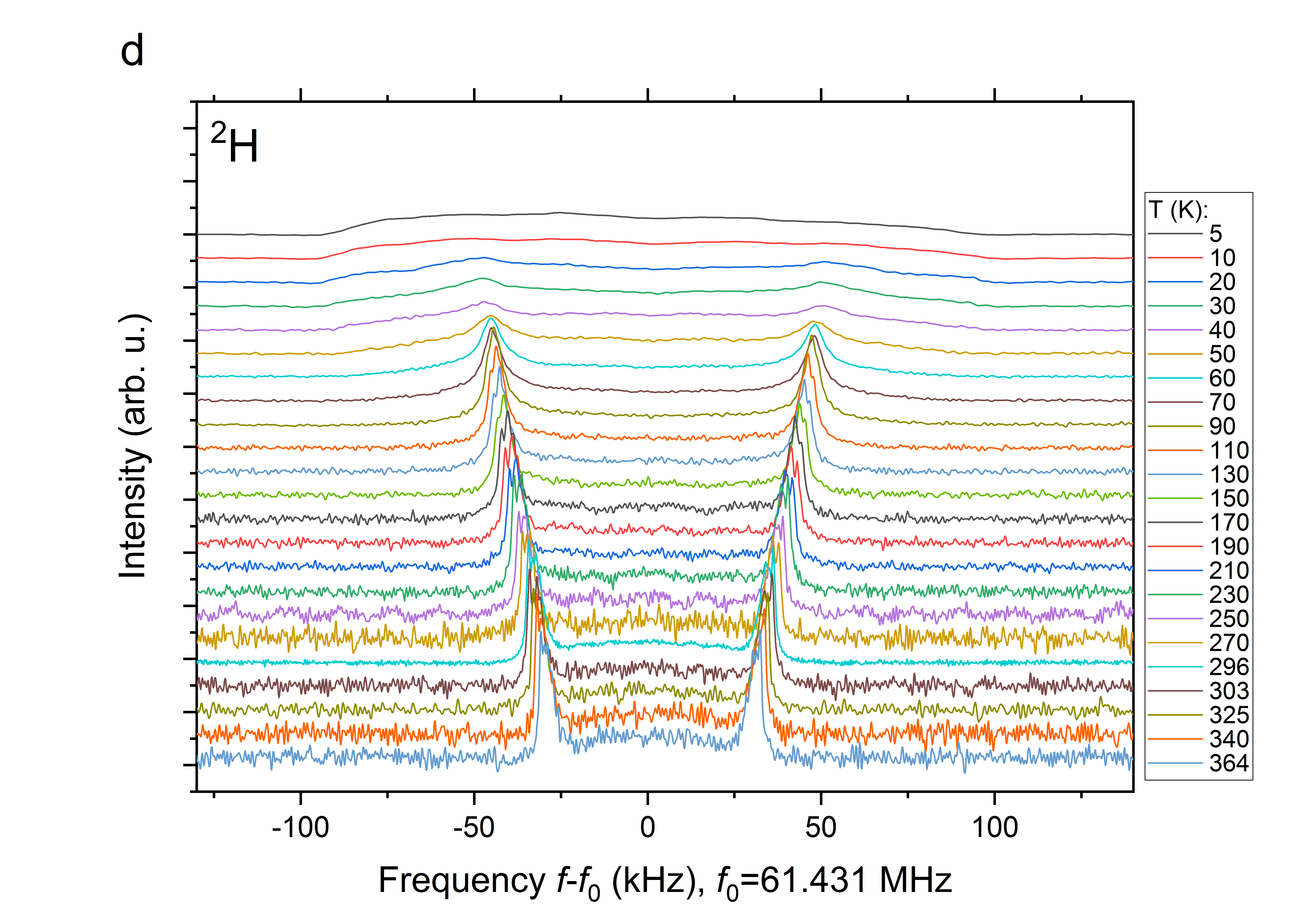}
\caption{\label{fig:TDSpec} Temperature dependences of \hyd and \deu NMR spectra of water in beryl single crystals. On top, \hyd spectra were recorded with the hexagonal axis of beryl sample oriented parallel (a) and perpendicular (b) with respect to the external magnetic field, respectively. Analogous configuration was used for \deu spectra displayed in the bottom (c, d).}
\end{figure}

The
temperature dependent \hyd and \deu NMR spectra within 5--\qty{364}{\kelvin} are
displayed in Fig.~\ref{fig:TDSpec}. The dipolar and quadrupole splittings
$\Delta_{\mathrm D}$, $\Delta_{\mathrm Q}$ increase upon cooling,
and they level off at about \qty{70}{\kelvin}. The magnitude of the observed \hyd
dipolar splitting and its temperature behavior are in agreement with \hyd
NMR data available in the literature \cite{Pare1964,Sugitani1966}, where this
behavior was explained by oscillations of the water molecules around
their equilibrium positions along the hexagonal axis. This
interpretation, however, conflicts with the fact that the dipolar
splittings do not reach the expected values even at the lowest
temperatures. This leads us unambiguously to the conclusion that the motions of the water molecules have to be more complex
than this. As for the \deu spectra, we are aware of no published measurements of the quadrupole splitting of \deu of \hw in beryl crystal. Our experiments reveal that the \deu spectra follow qualitatively the same pattern as the \hyd ones---the splittings of the quadrupole doublets decrease upon cooling in a similar manner.

Below \qty{70}{\kelvin}, the \hyd and \deu spectral peaks become significantly broader, indicating a slowing down of the molecular motions so that the averaging of the anisotropic interactions (magnetic dipolar for \hyd and electric quadrupole for \deu) becomes inefficient. Thus, at the lowest temperatures, the spectral shapes resemble features typical of static NMR spectra of randomly oriented powder samples. The temperature dependences of the measured NMR spectra and their spectral shapes are analyzed in more detail below.

\section{Analysis and Discussion}
\label{analysis}

In this section we show that the established notion of the behavior of water molecules in beryl, i.e., motions with equilibrium positions along the hexagonal axis, is not consistent with the measured NMR data.
Furthermore, we propose a new type of movement when the water molecule is bonded via a hydrogen bond to an oxygen atom in the enclosing wall of the beryl void. A detailed analysis allows us to explain all the features observed in the NMR experiment.

The key to determining the motions of the water molecule in beryl lies in a fundamental difference between the \hyd and \deu NMR spectra. The dipolar splitting of \hyd peaks depends on the orientation of the
\hyd--\hyd line with respect to the direction of external magnetic field (via
the angle $\vartheta_\mathrm{D}$ in Eq.~\ref{eq:dipolar}). In
contrast, the quadrupole splitting of \deu peaks depends on the direction
of the external magnetic field within the PAS of the EFG tensor (via the angles $\vartheta_\mathrm{Q}$ and $\varphi_\mathrm{Q}$ in Eq.~\ref{eq:quad}). The hydrogen atoms in the water molecule are (chemically) equivalent, and therefore, the \deu EFG tensors are of the same magnitude. However, the orientation of the PAS differs for the two \deu positions in the water molecule since for both EFG tensors the principal axis \VV points approximately along the respective O--H bond, and the water molecule is not linear (Fig.~\ref{fig:geom}a). As a consequence, the angles $\vartheta_\mathrm{Q}$ and $\varphi_\mathrm{Q}$ are in general different for the two \deu sites.

\subsection{Model for Motion of Water Molecules in Beryl Voids}
For sake of spectra analysis, let us provisionally assume that the type-I water molecules perform the traditionally accepted type of motion within the void of the beryl crystal as follows. Let the equilibrium orientation of the molecules be such that the H--H line is aligned with the hexagonal axis $\mathbf{c}$ of the beryl structure, and the molecules may rotate and oscillate around the H--H line within a limited spatial angle around axis $\mathbf{c}$---as dictated by the shape and dimensions of the void. Owing to such a motion, the angle $\vartheta_\mathrm{D}$ is averaged to zero for the case when the external field $\mathbf{B}_0\parallel \mathbf{c}$, or to $90^\circ$ for $\mathbf{B}_0 \perp \mathbf{c}$, and a qualitative agreement is reached with the temperature dependences of \hyd NMR (Fig.~\ref{fig:TDSpec}) and with \hyd NMR data in previous works.\cite{Pare1964,Sugitani1966} However, the observed dipolar splittings $\Delta_{\mathrm D}$ are somewhat lower than expected based on the distance of \hyd nuclei in the water molecule, suggesting that the angle $\vartheta_\mathrm{D}$ is on average different from 0 or $90^\circ$. An even more convincing argument against the considered type of motion is revealed by \deu NMR where two distinct quadrupole doublets are observed in the spectra for $\mathbf{B}_0 \parallel \mathbf{c}$ (Fig.~\ref{fig:spectraRT}b). Because of the symmetry of the water molecules, any rotational or oscillatory motion about the H--H axis would average the angle $\vartheta_\mathrm{Q}$ to the same value for both \deu nuclei, and hence the quadrupole splitting of \deu would be the same. But it is clear from the measured \deu NMR spectra that the quadrupole splitting, and thus also the averaged angle $\vartheta_\mathrm{Q}$, is significantly different for each of the two sites of deuterium in the water molecule.

\begin{figure}
\hfill		\subfloat{\includegraphics[width=0.35\columnwidth,valign=t]{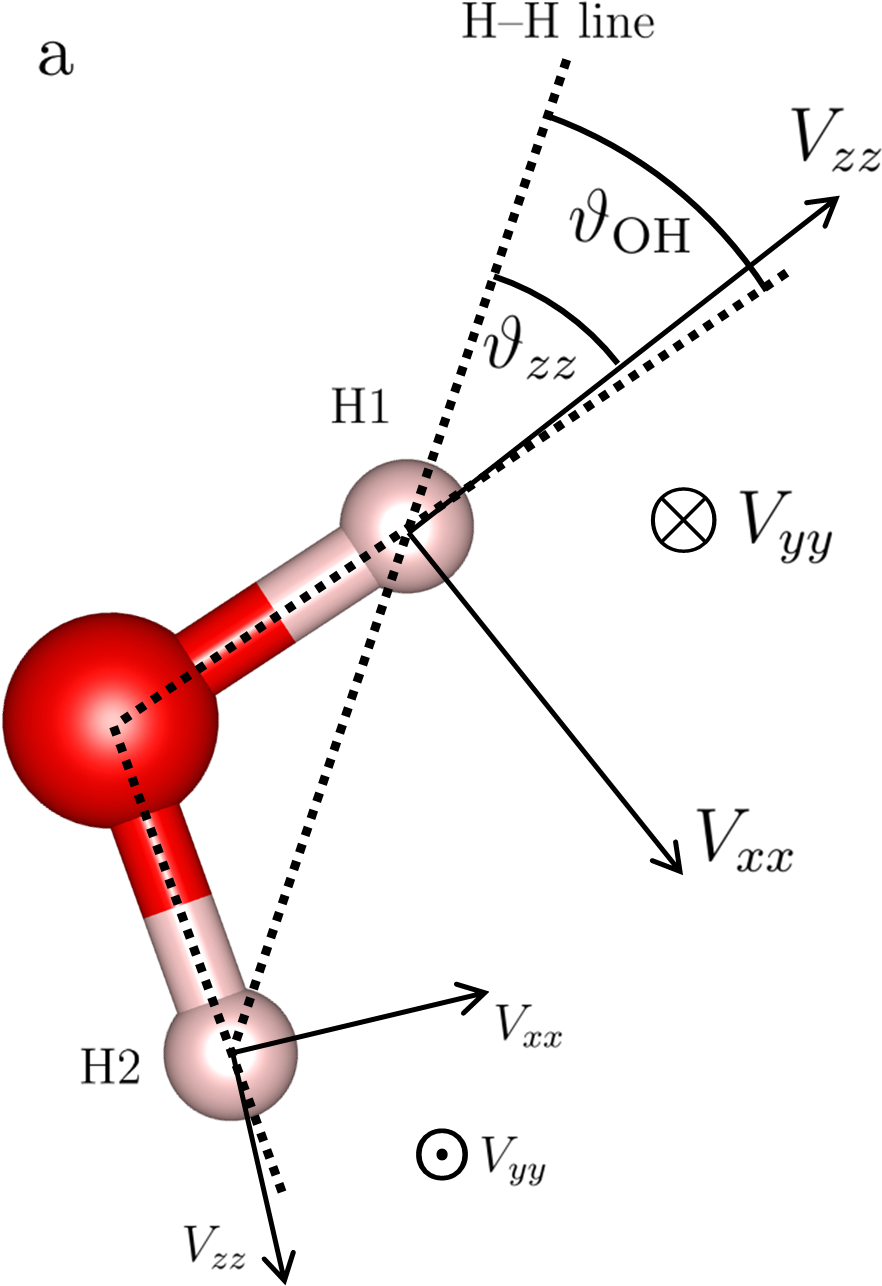}}\hfill
		\subfloat{\includegraphics[width=0.35\columnwidth,valign=t]{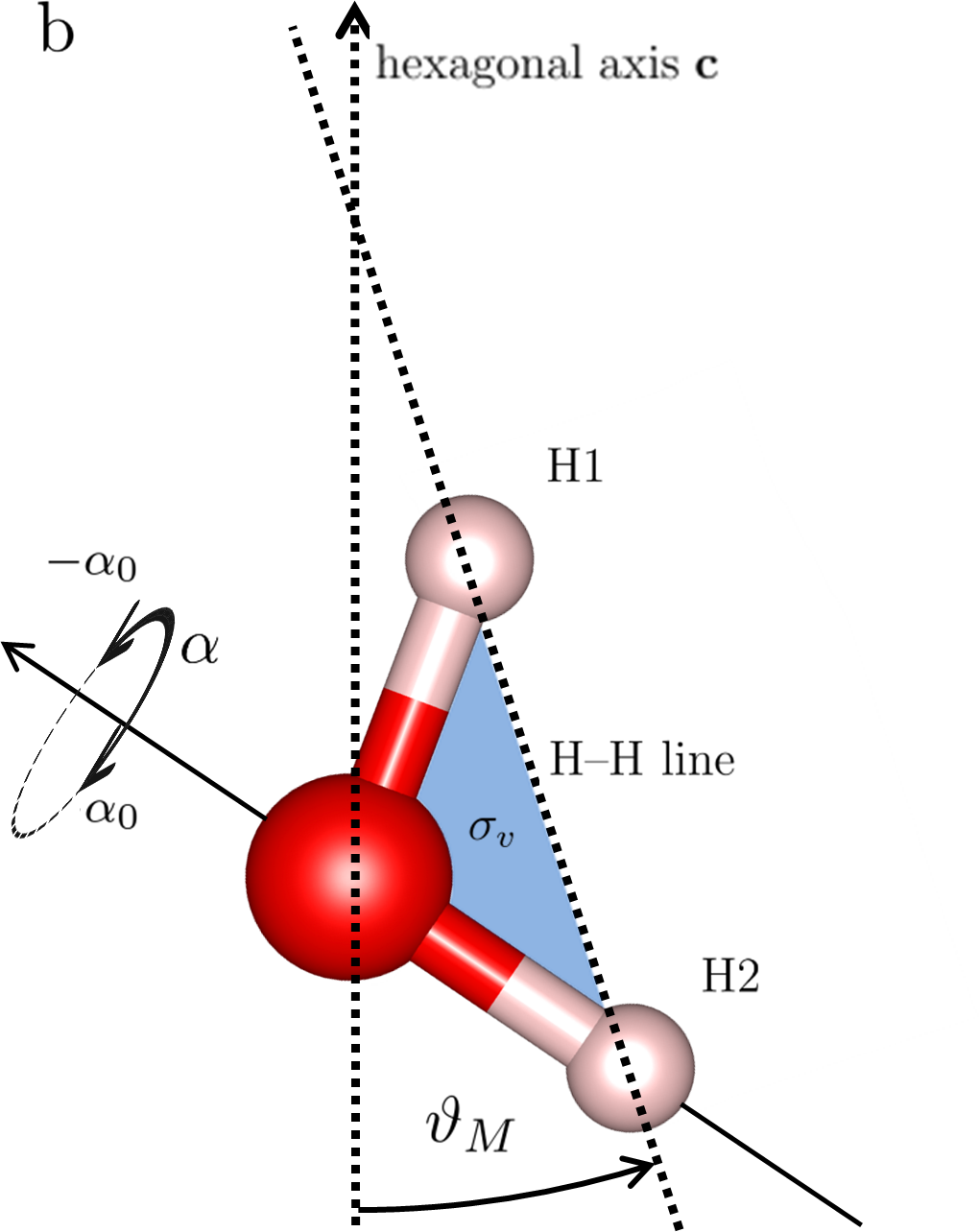}}\hfill \\
\hfill		\subfloat{\includegraphics[width=0.35\columnwidth]{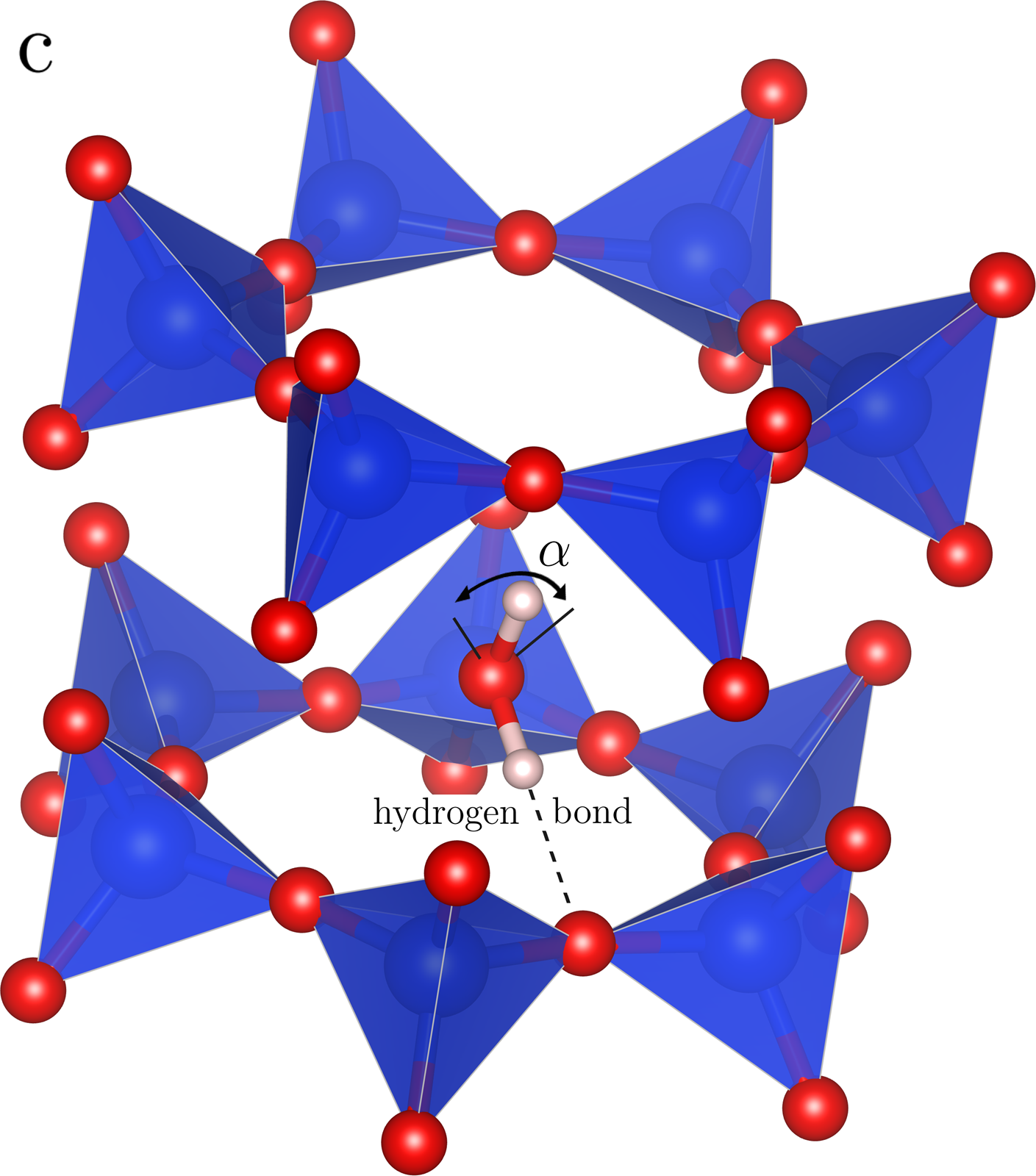}} \hfill
		\subfloat{\includegraphics[width=0.35\columnwidth]{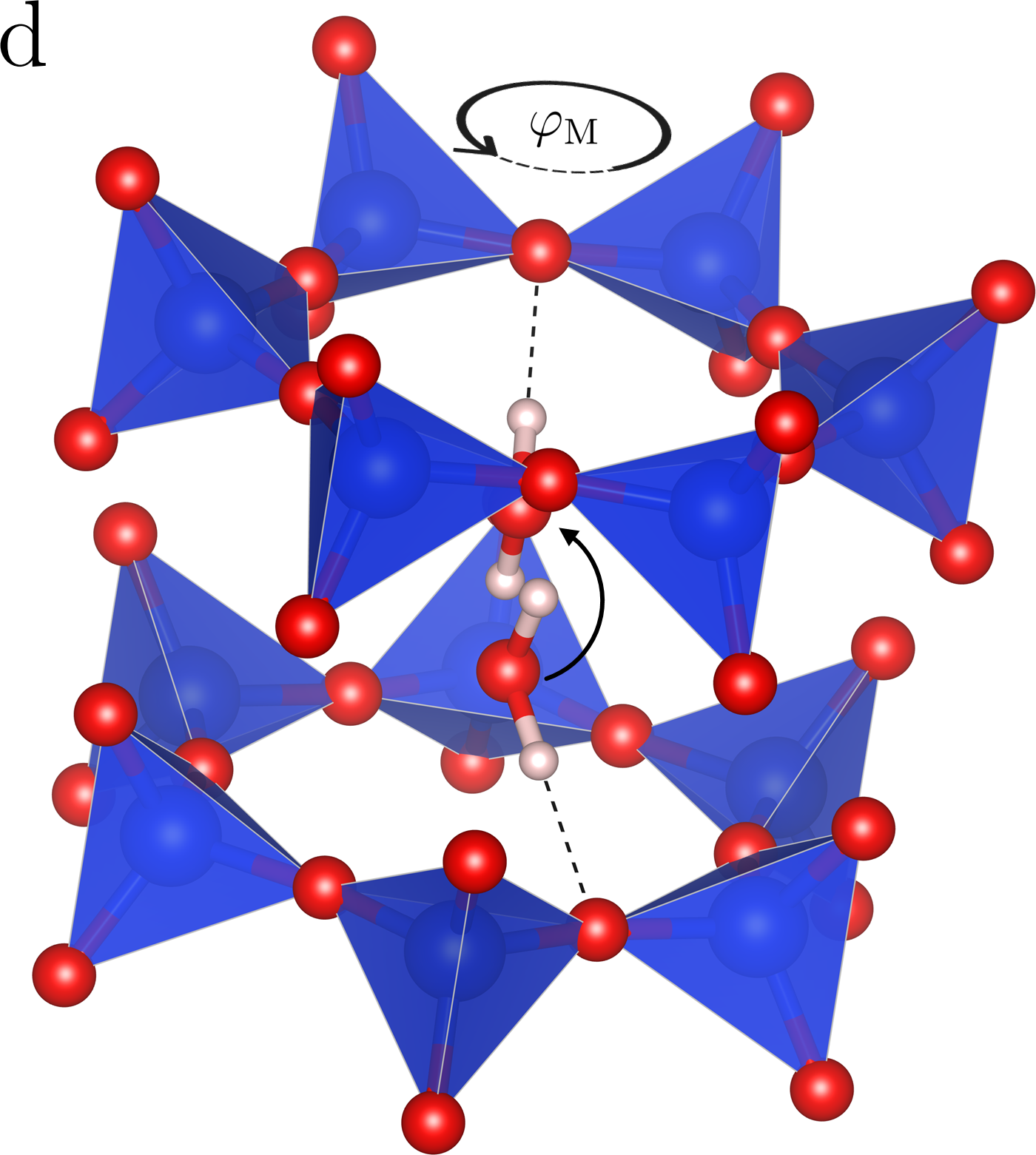}} \hfill
		\caption{\label{fig:geom} (a) The principal axis system of the EFG tensor at the hydrogen sites of the water molecule. Definition of the angles $\vartheta_{zz}$ and $\vartheta_{\mathrm{OH}}$ is shown. (b) Orientation of the water molecule with respect to the hexagonal axis. The angle $\vartheta_{M}$ defines the deviation of the H--H line from the hexagonal axis, the angle $\alpha$ describes librations around the axis O--H2 within range $\alpha\in\left(-\alpha_0,\alpha_0\right)$. (c) Proposed librational movement of the water molecule in the beryl void. The axis of libration (defined by the O--H2 bond) is aligned with the hydrogen bond which is formed between the H2 and one of the twelve oxygen atoms in the walls. (d) Proposed molecular jumps among different bonding sites in the void lead to effective rotations around the hexagonal axis (angle $\varphi_\mathrm{M}$).}
\end{figure}

In principle, the large quadrupole splitting of the outer doublet in the \deu spectrum could be compatible with a situation where the H--H line lies approximately along the hexagonal axis, but the splitting of the second, inner doublet would not comply. The quadrupole splitting of the inner doublets is so small that it can only be reached if the magnetic field maintains a very specific orientation with respect to the PAS of EFG of such a \deu atom: the angle $\vartheta_\mathrm{Q}$ must be (on average) close to the value  $54.7^\circ$, so called "magic angle" for which the expression $3\cos^2\vartheta_\mathrm{Q}\!-\!1$ in Eq.~\ref{eq:quad} becomes zero. This condition can be satisfied when the H--H line of the water molecule is deviated from the beryl hexagonal axis by an angle $\vartheta_\mathrm{M}\!\approx\!19.5^\circ$ (Fig.~\ref{fig:geom}b).  Since the angle $\vartheta_{zz}$ between the principal axis \VV of the EFG tensor and
the H--H line is $35.2^\circ$, then
$\vartheta_\mathrm{Q}=\vartheta_{zz}\pm\vartheta_\mathrm{M}$, yielding the values
of $54.7^\circ$ and $15.7^\circ$, the first one being close to the magic angle. 
 
The reason for such peculiar arrangement of the water molecules is apparently the
formation of a hydrogen bond between the deuterium atom in \hw and one of the oxygen atoms which form the walls of the beryl voids. Therefore, we propose a model in which the water molecules attach to an oxygen atom via a hydrogen bond, and the molecules perform two distinct motions: first, they librate rapidly about the formed hydrogen bond (Fig.~\ref{fig:geom}c). Since hydrogen bonds tend to be highly directional,\cite{Wood2009} the axes of libration coincide with the O--H bonds of the molecules, and these axes deviate from the beryl hexagonal axis $\mathbf{c}$ by an angle close to the magic angle $54.7^\circ$. Second, since
the water molecules may form the hydrogen bonds with any of
the twelve oxygen atoms available in the walls of the voids, the
molecules are also expected to jump among these available bonding sites  (Fig.~\ref{fig:geom}d), i.e.,
they effectively rotate about the beryl hexagonal axis. In the following analysis, we show that these two types of motion can explain the NMR experiment
even quantitatively.

Naturally, the jumps accompanied by re-bonding of hydrogen bond have to be
slower than the librations about the O--H bond, yet both these molecular motions
are fast enough to cause averaging effects in NMR. The resonance
frequencies of the nuclei are usually orders of magnitude slower than the
changes of local fields induced by such molecular motions. As a consequence, the
measured \hyd and \deu NMR spectra of water in beryl are affected by averaged
dipolar and quadrupole interactions, which is documented by the reduced
values of splittings and by the presence of narrow spectral peaks at
higher temperatures. Upon lowering the temperature, the frequencies and
amplitudes of molecular motions decrease, therefore the splittings
approach their nominal values, and the spectral peaks broaden. 

\subsection{Analysis of Librations of Water Molecules}

\begin{figure}
	\includegraphics[width=0.5\columnwidth]{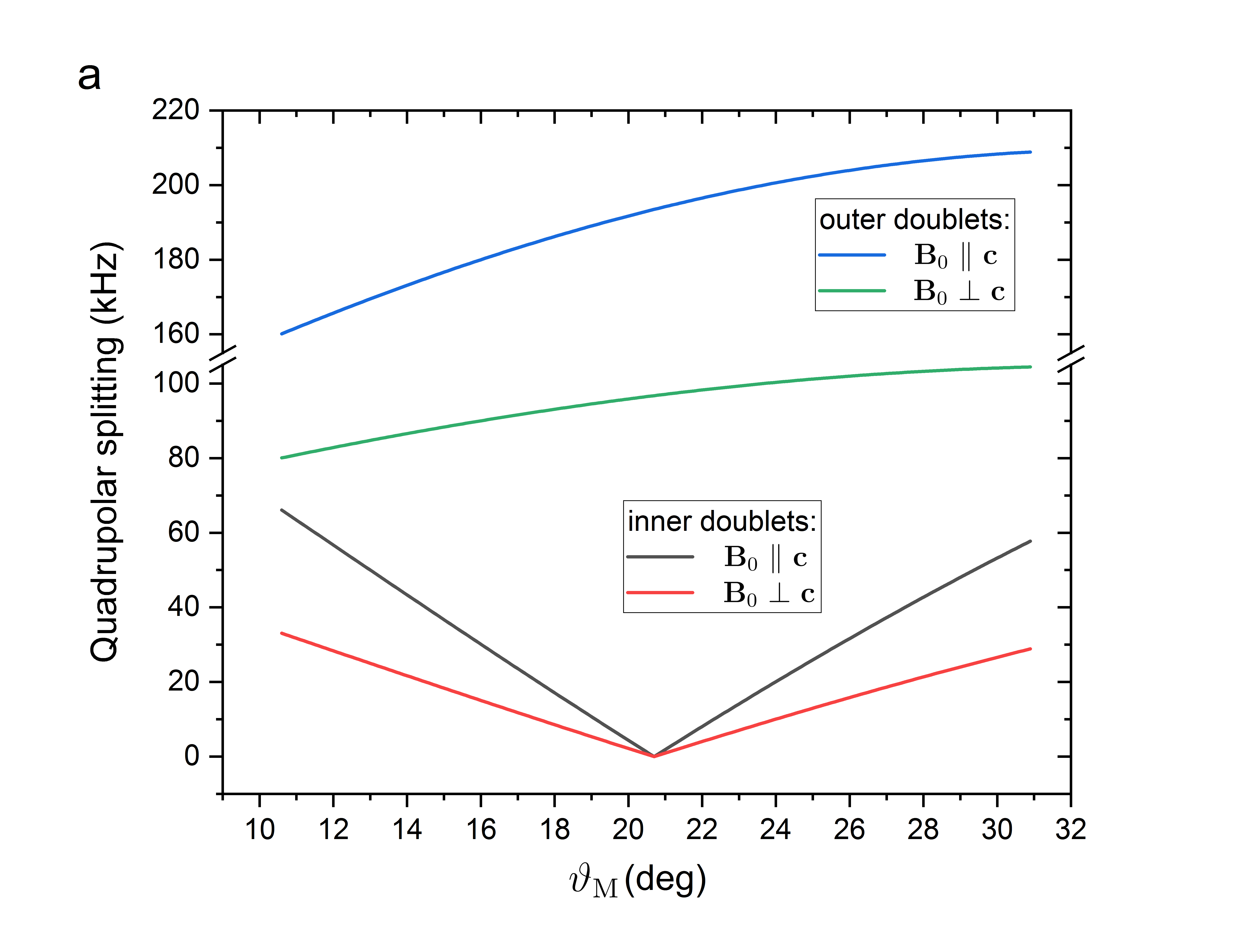}\includegraphics[width=0.5\columnwidth]{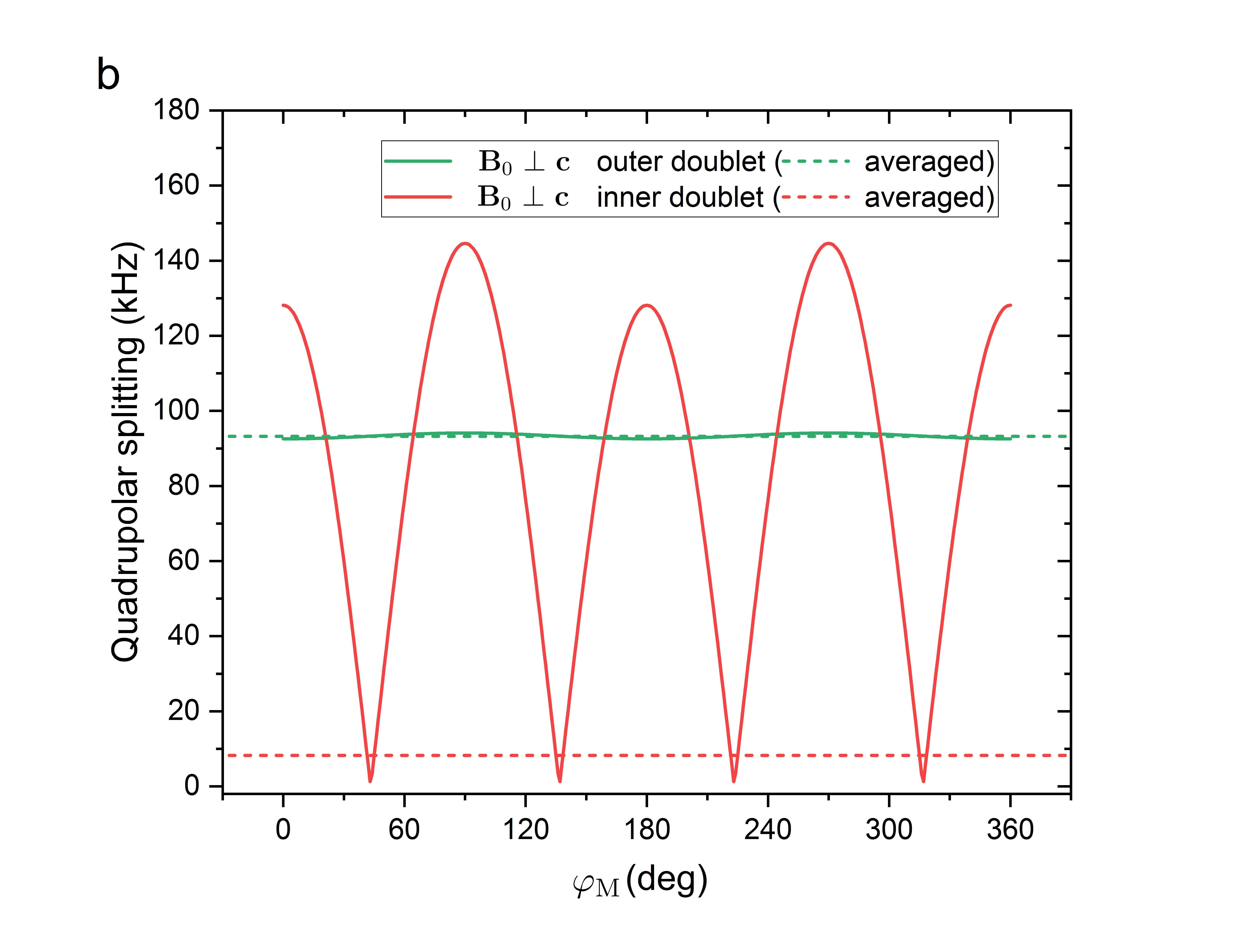}
	\caption{\label{fig:sensitivity} (a) Dependence of the
		quadrupole splittings $\left<\left<\Delta_\mathrm{Q}\right>\right>_{\mathrm{OH},c}$ of doublets in \deu spectra on the tilt angle
		$\vartheta_\mathrm{M}$. The splittings were averaged by librations around O--H bond (amplitude $\alpha_0 = 40^\circ$) and jumps around hexagonal axis (averaging with respect to the angle $\varphi_\mathrm{M}$).  (b) Sensitivity of the quadrupole splittings $\left<\Delta_\mathrm{Q}\right>_{c}$  to the angle $\varphi_\mathrm{M}$ in case of perpendicular orientation of the magnetic field, angle $\vartheta_\mathrm{M}=18.1^\circ$, and averaging only by the librations ($\alpha_0 = 40^\circ$). Dashed lines correspond to values when the splittings were additionally averaged with respect to the angle $\varphi_\mathrm{M}$.}
\end{figure}

\begin{figure}
	\includegraphics[width=0.5\columnwidth]{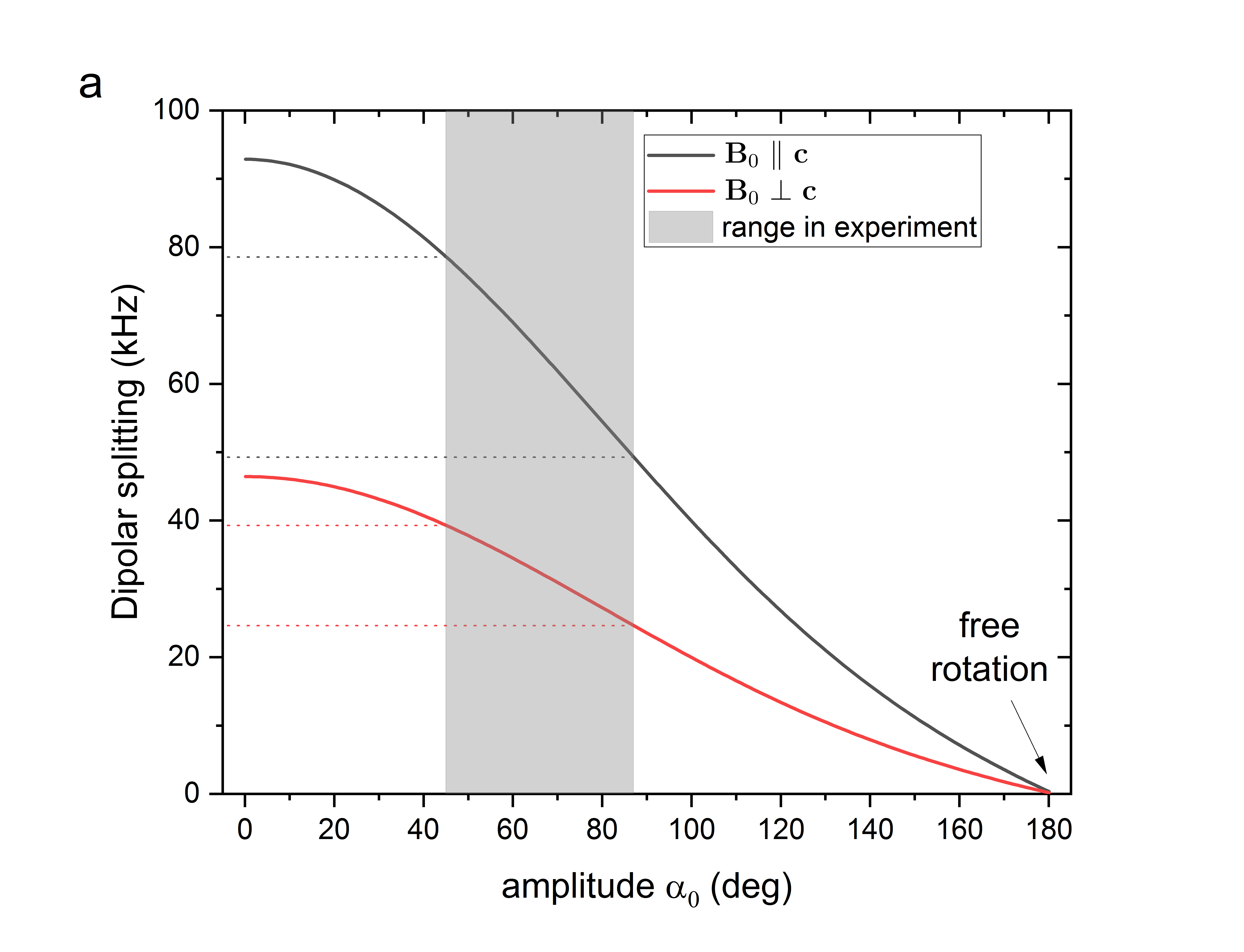}\includegraphics[width=0.5\columnwidth]{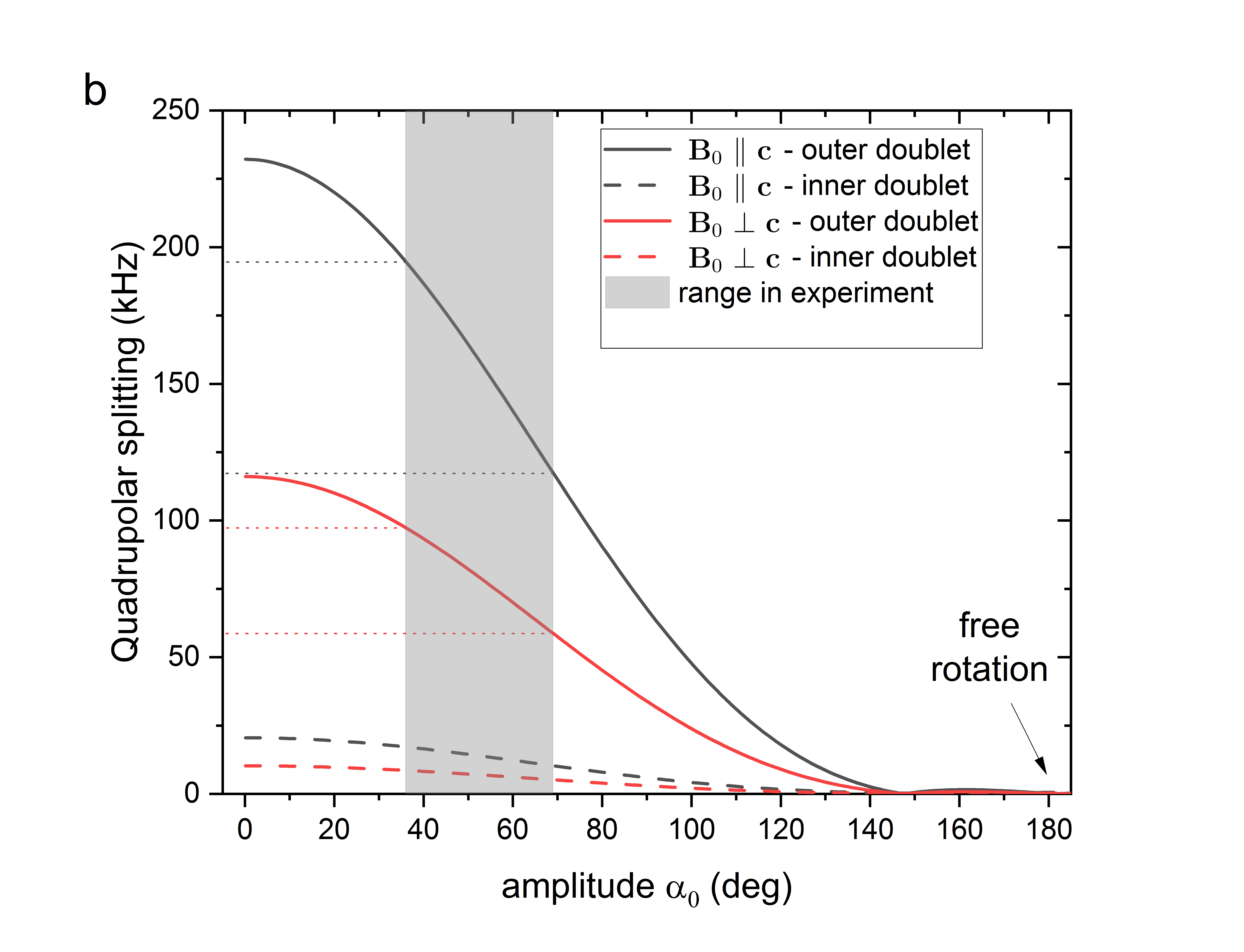}
	\caption{\label{fig:alphadep} Calculated averaged dipolar (a) and quadrupole (b) splitting as a function of the angular amplitude of librations $\alpha_0$. The water molecule is assumed to rotate uniformly within an angular interval  $\left(-\alpha_0,\alpha_0\right)$ about its O--H2 axis tilted by $\vartheta_\mathrm{M}\!\approx\!18.1^\circ$. Additionally the molecule jumps between different bonding sites (averaging with respect to angle $\varphi_\mathrm{M}$). Grey areas denote the range of observed values $\alpha\_0$.}
\end{figure}

\begin{figure}
	\includegraphics[width=0.5\columnwidth]{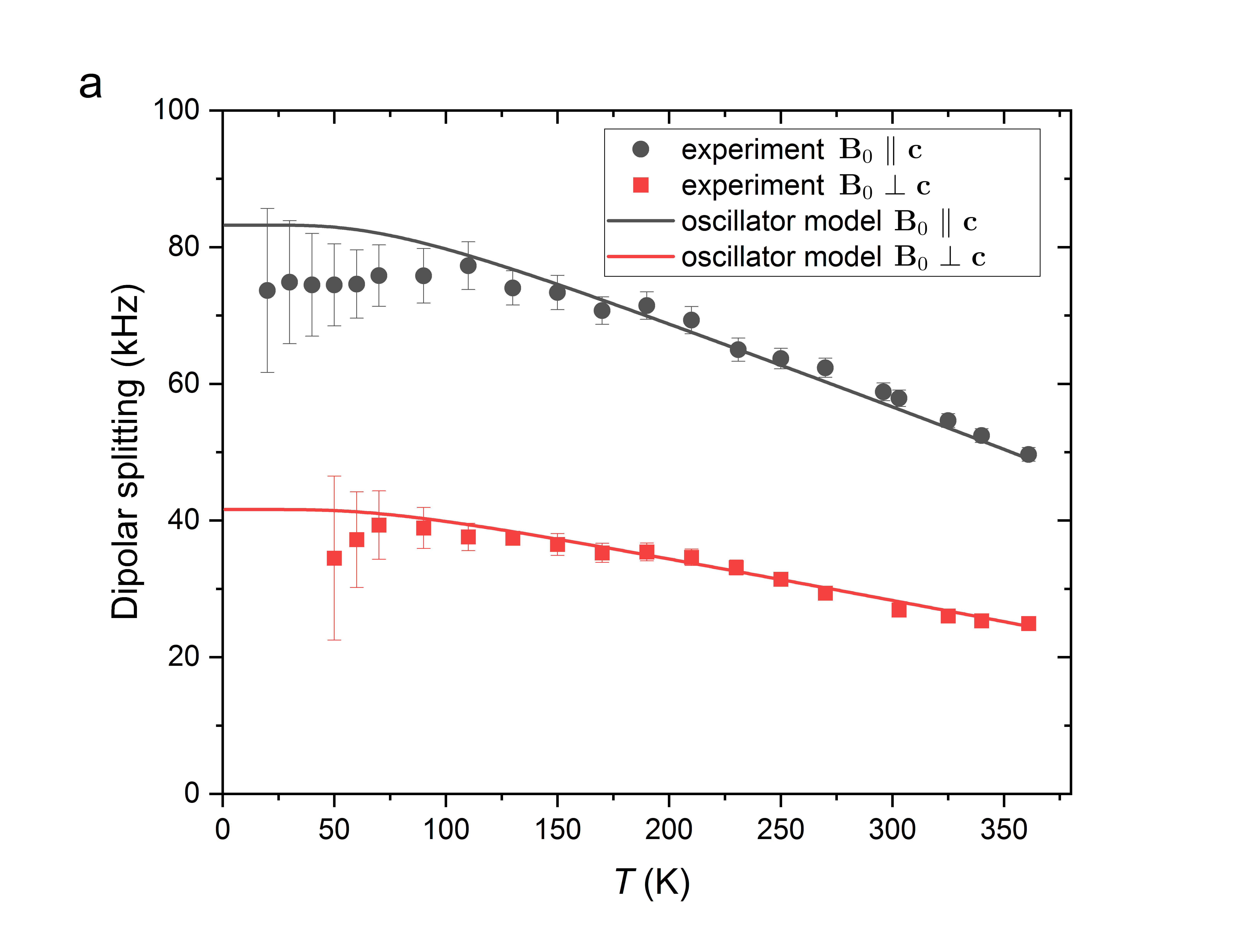}\includegraphics[width=0.5\columnwidth]{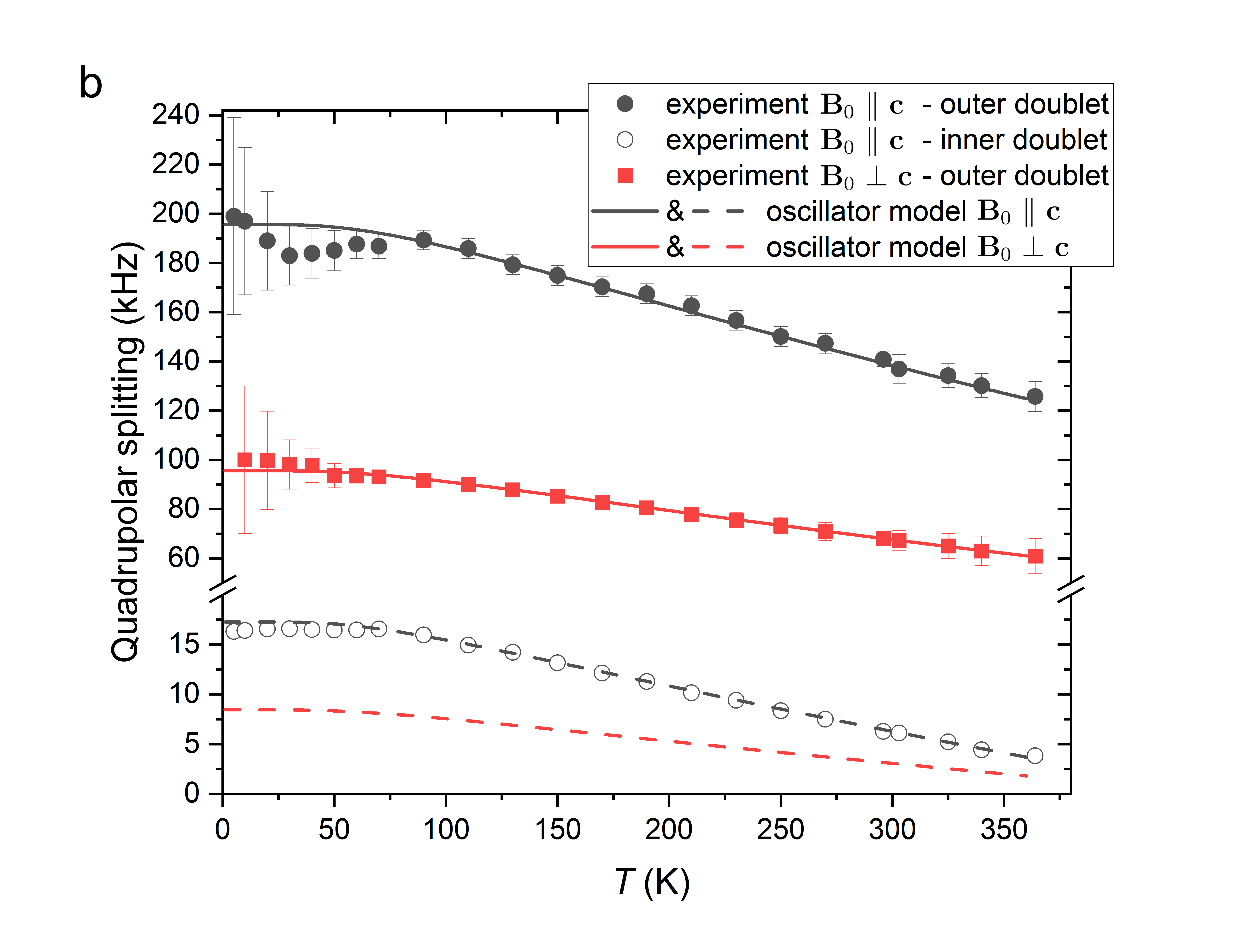}
	\caption{\label{fig:compare} Comparison of dipolar (a) and quadrupole (b) splittings measured as a function of temperature with those obtained by the model based on harmonic oscillator.}
\end{figure}

In order to analyze the first mode of molecular motion (i.e., the rapid
librations of the molecule about the hydrogen bond), we denote by $\vartheta_\mathrm{M}$ the angle
describing the deviation of the H--H line from the hexagonal axis of the beryl crystal
(Fig.~\ref{fig:geom}b). The librations of the molecule about its axis
O--H2 are
described by the angle $\alpha$ which is zero for the apex position (i.e., when the hexagonal axis $\mathbf{c}$ is parallel to the mirror plane $\sigma_v$ which contains all three atoms of the water molecule), and
we suppose it librates around the mean position
with an amplitude of $\alpha_0$, $\alpha\in\left(-\alpha_0,\alpha_0\right)$. For now we assume the angle $\alpha$ to be distributed uniformly within its limits
$\left(-\alpha_0,\alpha_0\right)$, which enables us to express the average dipolar and quadrupole splittings as:

\begin{eqnarray}
	\left<\Delta_\mathrm{D}\right>_\mathrm{OH} &=&\frac{1}{2\alpha_0}\int_{-\alpha_0}^{\alpha_0} \Delta_\mathrm{D}\, \mathrm{d}{\alpha}, \nonumber  \\
	\left<\Delta_\mathrm{Q}\right>_\mathrm{OH} &=&\frac{1}{2\alpha_0}\int_{-\alpha_0}^{\alpha_0} \Delta_\mathrm{Q}\, \mathrm{d}{\alpha}	\label{eq:average1}
\end{eqnarray}

The second type of motion is connected with re-bonding of the H2 atom to any of
the remaining five oxygen sites or, instead, with H1 atom forming a hydrogen
bond with one of the six oxygen sites in the other hemisphere (Fig.~\ref{fig:geom}d). Such jumps
cause averaging with respect to the hexagonal axis $\mathbf{c}$ of beryl, which we
can consider as rotational averaging over the angle $\varphi_\mathrm{M}$ in the full $\left<0,2\pi\right)$ range. The dipolar and quadrupole splittings, averaged by both types of movements, can then be expressed as functions of the angles
$\vartheta_\mathrm{OH}$, $\vartheta_{zz}$, $\vartheta_\mathrm{M}$, and
$\alpha_0$:

\begin{eqnarray}
	\left<\left<\Delta_\mathrm{D}\right>\right>_{\mathrm{OH},c} &=&\frac{1}{2\pi}\frac{1}{2\alpha_0}\int_{0}^{2\pi}\int_{-\alpha_0}^{\alpha_0} \Delta_\mathrm{D}\, \mathrm{d}{\alpha}\, \mathrm{d}{\varphi_\mathrm{M}} \nonumber \\
	\left<\left<\Delta_\mathrm{Q}\right>\right>_{\mathrm{OH},c} &=&\frac{1}{2\pi}\frac{1}{2\alpha_0}\int_{0}^{2\pi}\int_{-\alpha_0}^{\alpha_0} \Delta_\mathrm{Q}\, \mathrm{d}{\alpha}\, \mathrm{d}{\varphi_\mathrm{M}} \label{eq:average2}
\end{eqnarray}

The averaged splittings
$\left<\left<\Delta_\mathrm{D}\right>\right>_{\mathrm{OH},c}$ and
$\left<\left<\Delta_\mathrm{Q}\right>\right>_{\mathrm{OH},c}$ further
depend also on the dipolar constant $\delta$ and the
quadrupolar parameters \cq and
$\eta$, respectively. The value of $\delta$ is given by the distance between
the \hyd nuclei (Eq.~\ref{eq:dipolar}) and thus it is closely linked to
the geometry of the water
molecule. \cq and $\eta$ are unknown for water confined in beryl but they should not
be dramatically different from the values for \hw in various phases,
i.e., \cq$\!\approx\!200$--\qty{300}{\kHz}
and $\eta\!\approx\!0.1$. Also the angles $\vartheta_\mathrm{OH}$ and
$\vartheta_{zz}$ are given by the geometry of water molecule
(Fig.~\ref{fig:geom}a). In further
analysis, we assume the value of the H--O--H angle of
$106.9^\circ$ and a H--O
distance of \qty{0.949}{\angstrom}, which was found by neutron diffraction experiments
\cite{Gatta2006,Kolesnikov2016}. These parameters yield a value of
$\vartheta_{\mathrm{OH}}=36.55^\circ$ for the angle between the H--H line and the
O--H bond. The angle $\vartheta_{zz}$ between the H--H line and the
principal axis \VV is then equal to $35.2^\circ$, which corresponds to a small
deviation ($\approx\!1.35^\circ$) of the \VV axis from the direction of
the O--H bond, as
determined from hyperfine structure measurements of heavy
water\cite{Bluyssen1967}. The H--H distance is equal to \qty{1.525}{\angstrom} and thus
$\delta\!\approx\,$\qty{58.7}{\kHz}, according to Eq.~\ref{eq:dipolar}. Based on these values, the remaining parameters of
interest, angles
$\vartheta_\mathrm{M}$ and $\alpha_0$, can be determined from our NMR experiments
using a suitable model.

The angle $\vartheta_\mathrm{M}$ determines the deviation of the
libration axis O--H from the hexagonal axis and its value dictates,
to a great extent, the specific character of the observed \deu spectra, i.e.,
the positions of the two doublets with distinctly different quadrupole splittings.
Whereas the dipolar splitting in \hyd spectra as well as the quadrupole
splitting of outer doublets in \deu spectra do not vary much with deviation of
the axis of libration (angle $\vartheta_\mathrm{M}$), the small quadrupole
splitting of the inner doublet in the \deu spectra is particularly
sensitive to the changes in the $\vartheta_\mathrm{M}$ angle
(Fig.~\ref{fig:sensitivity}a). For both sample orientations
$\mathbf{B}_0\perp\mathbf{c}$ and $\mathbf{B}_0\parallel\mathbf{c}$, the
splitting of the inner doublet becomes zero when
$\vartheta_\mathrm{Q}=\vartheta_{zz}+\vartheta_\mathrm{M}$ reaches the magic
angle. One may notice that there are two possible values of
$\vartheta_\mathrm{M}$ that yield the small splitting of the inner
doublets observed in experiment; this issue will be addressed in further analysis
below. 

The peaks of the inner doublet in the \deu spectrum become
substantially broadened
for $\mathbf{B}_0\perp\mathbf{c}$ (as well as for any orientations where the
external magnetic field $\mathbf{B}_0$ is deviated from the hexagonal axis
$\mathbf{c}$ by more than $\approx\!30^\circ$, see Fig.~\ref{fig:angular}), and
this broadening becomes even more pronounced with decreasing temperature.
For $\mathbf{B}_0\perp\mathbf{c}$, the splitting of the inner doublet
depends strongly on the angle $\varphi_\mathrm{M}$ (Fig.~\ref{fig:sensitivity}b). Consequently, for the inner doublet, averaging over the angle $\varphi_\mathrm{M}$
becomes ineffective already at temperatures below room
temperature, whereas the averaging is still effective enough for the outer
doublet to be well resolved. The averaging of $\left<\left<\Delta_\mathrm{D}\right>\right>_{\mathrm{OH},c}$ and $\left<\left<\Delta_\mathrm{Q}\right>\right>_{\mathrm{OH},c}$ in Eq.~\ref{eq:average2} is assumed to be continuous in angle $\varphi_\mathrm{M}$, whereas in reality there are twelve discrete orientations in the single crystal, i.e., the angle $\varphi_\mathrm{M}$ should take values $\frac{\pi}{6}k+\varphi_\mathrm{M,0}$, where $k\in\left<0,12\right)$ is an integer. However, neither this simplification nor the indeterminacy of the phase angle $\varphi_\mathrm{M,0}$ lead to any noticeable difference in the averaged values of the splittings.

Although both $\vartheta_\mathrm{M}$ and $\alpha_0$ are expected to be
temperature dependent, mainly the amplitude of librations $\alpha_0$ is
responsible for the decrease in dipolar and quadrupole splittings with
temperature due to motional averaging. The averaged splittings
$\left<\left<\Delta_\mathrm{D}\right>\right>_{\mathrm{OH},c}$ and
$\left<\left<\Delta_\mathrm{Q}\right>\right>_{\mathrm{OH},c}$ as functions of
amplitude $\alpha_0$ are displayed in Fig.~\ref{fig:alphadep}; the grey
areas denote the ranges of amplitudes corresponding to  experimentally observed dipolar and quadrupole splittings. The splittings obtained by NMR experiments can thus be related to variations of the amplitude $\alpha_0$ within the given range.
But even if the calculated dependences roughly capture the range and basic
character of experimental splittings (lower values of $\alpha_0$ correspond to
low temperatures and highest $\alpha_0$ to $T\!\!\approx\,$\qty{360}{\kelvin}), it is clear that
there are systematic differences. First, the experimentally detected
low-temperature plateaus are not present in the simulated dependences. Second, in the experiment the splitting of the inner quadrupole doublet decays faster
with increasing temperature than the splitting of outer doublet---the ratio of
splittings for these doublets gradually increases from 11.5 at low temperatures
to more than 30 at high temperatures (Fig.~\ref{fig:compare}).

Both these shortcomings can be removed by implementing a specific temperature
dependence to angles $\vartheta_\mathrm{M}$ and $\alpha_0$. Such dependence,
however, may be realized in many different ways leading to similar levels
of agreement
with the experiment. We show that a good agreement (Fig.~\ref{fig:compare}) is
reached already with probably the simplest approach when the librational
motion of the water molecule about one of its O--H axes is considered as
that of a quantum harmonic oscillator.

A quantum harmonic oscillator provides quantized energies
\begin{eqnarray}
E_n=\hbar \omega (n + \frac{1}{2}),	
	\label{eq:osc_ene}
\end{eqnarray}
where $\omega$ denotes the angular frequency of the librational motions of the water molecule about the O--H bond. The states with energies $E_n$ are populated with Boltzmann probabilities
\begin{eqnarray}
	p_n= \frac{e^{-\frac{E_n}{k_B T}}}{\sum_{n=0}^{\infty}e^{-\frac{E_n}{k_B T}}}, 	\label{eq:osc_prob}
\end{eqnarray}
the mean energy can be written as
\begin{eqnarray}
	\left< E \right>= \frac{\sum_{n=0}^{\infty}E_n e^{-\frac{E_n}{k_B T}}}{\sum_{n=0}^{\infty}e^{-\frac{E_n}{k_B T}}} 	\label{eq:osc_mean}
\end{eqnarray}
and it is related to the amplitude $\alpha_0$ as 
\begin{eqnarray}
	\left< E \right>= \frac{1}{2} k\left<y^2\right> = \frac{1}{2}  k d_\mathrm{OH}^2 \left<   \sin^2 \alpha_0 \right>, 	\label{eq:osc_ampl}
\end{eqnarray}
where $d_\mathrm{OH}$ is the length of the O--H bond. The force
constant $k$ of the oscillator is related to the frequency $\omega$ as
$k=m\omega^2$, where $m$ is the mass of proton or deuteron for the \w or \hw case,
respectively. The value of $\omega$ is a free parameter in the harmonic
oscillator model, and it accounts for the overall shape of the temperature
dependences.

Eqs.~\ref{eq:average1} and \ref{eq:average2}, which assumed a uniformly
distributed angle $\alpha$, now have to be modified because for a
Boltzmann-averaged harmonic oscillator the angle $\alpha$ is not distributed
uniformly; instead, it has a Gaussian distribution\cite{Pathria2011}
with a dispersion $\alpha_0^2$. In our case only a few lowest energy levels of the oscillator are populated even at the highest temperature $T\!=\,$\qty{365}{\kelvin}, therefore, the zero-point energy is considerable and a visible plateau up to about \qty{50}{\kelvin} appears in all simulated dependences (Fig.~\ref{fig:compare}).  

The fact that upon heating, the splitting of the inner quadrupole doublet decays faster than that of the outer doublet can be captured
by making the angle $\vartheta_\mathrm{M}$ temperature dependent. By increasing
$\vartheta_\mathrm{M}$  slightly with increasing temperature, the value of
$\vartheta_\mathrm{Q} =\vartheta_{zz} + \vartheta_\mathrm{M}$ further
approaches the magic angle, which reduces the splittings of the inner
quadrupole doublets while leaving the splittings of the outer doublets relatively
unchanged (see Fig.~\ref{fig:sensitivity}a). If we thus increase
$\vartheta_\mathrm{M}$ within the range of 18.1--19.2$^\circ$ linearly with temperature in
range 5--\qty{365}{\kelvin}, the ratio of simulated splittings follows very well the
values observed in our experiment (Fig.~\ref{fig:compare}).

\begin{figure}	
	\includegraphics[width=0.5\columnwidth]{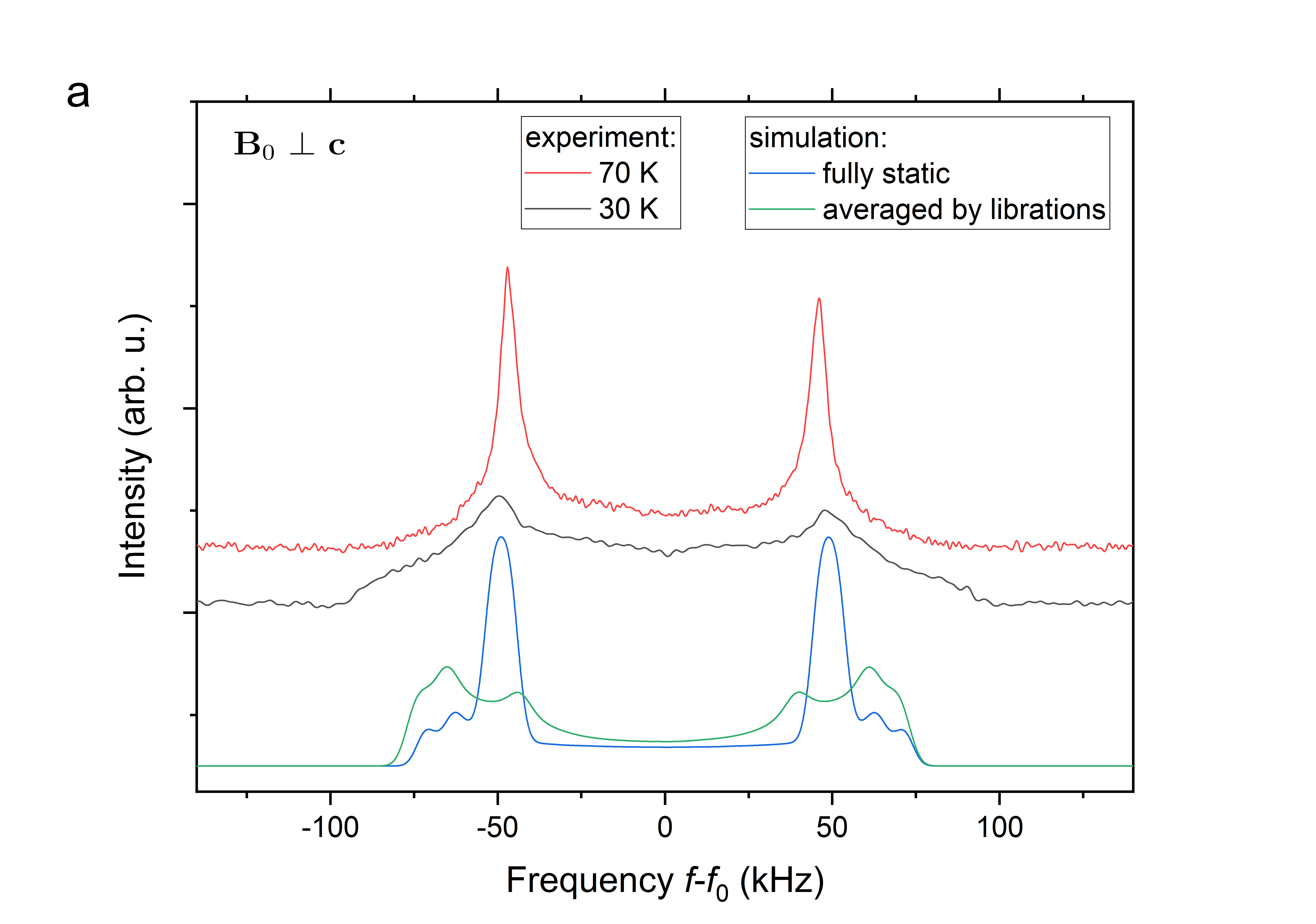}\includegraphics[width=0.5\columnwidth]{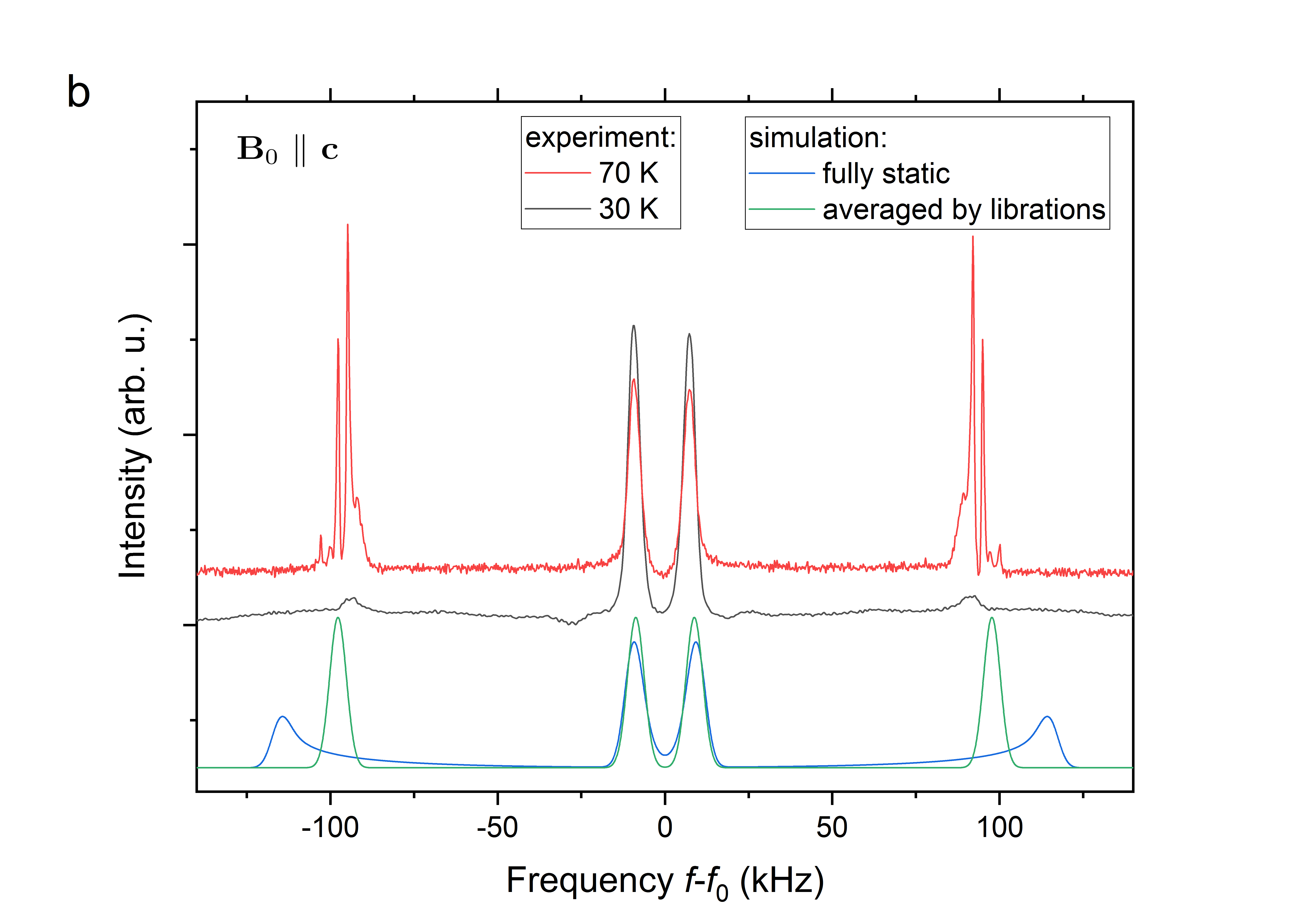}	
	\caption{\label{fig:lowT} Experimental \deu NMR spectra of water in beryl at low temperatures for $\mathbf{B}_0 \perp \mathbf{c}$ (a) and $\mathbf{B}_0 \parallel \mathbf{c}$ (b). Simulated NMR spectra with powder-like features were construced by removing the molecular jumps (averaging via angle $\varphi_\mathrm{M}$) or by removing the motions entirely, i.e. considering static, randomly halted molecules.}
\end{figure}

In the dependence of quadrupole splittings on angle $\vartheta_\mathrm{M}$
(Fig.~\ref{fig:sensitivity}a), there are apparently two values of
$\vartheta_\mathrm{M}\!\approx\!18^\circ$ and $\approx\!23^\circ$, bringing
$\vartheta_\mathrm{Q}$ close to the magic angle and thus leading to the small
splittings of inner doublets in the experiment. The larger value of
$\vartheta_\mathrm{M} $ can also lead to a good agreement of the model with
experimental quadrupole splittings, using a different frequency $\omega$, similar
quadrupole parameters \cq and $\eta$, and a gradual decrease in
$\vartheta_\mathrm{M}$ with increasing temperature---again to approach the
magic angle for $\vartheta_\mathrm{Q}$, but this time from above. To
discriminate between these two solutions we can use the dipolar splittings in
\hyd spectra, where this somewhat larger tilt angle $\vartheta_\mathrm{M}\!\approx\!23^\circ$ does
not yield a reasonable agreement. In order to match the experiment, we have to use
$\delta\!\approx\,$\qty{69.1}{\kHz} corresponding to a H--H distance of
\qty{1.444}{\angstrom}, which would lead to an unrealistic H--O--H angle of
$99.1^\circ$ (for an O--H
bond-length of \qty{0.949}{\angstrom}). Moreover, a decrease in the tilt angle $\vartheta_\mathrm{M}$
with increasing temperature would be required, which lacks a suitable
justification. In contrast, for the $\vartheta_\mathrm{M}\!\approx\!18^\circ$
solution, an increase in $\vartheta_\mathrm{M}$ with temperature makes more
physical sense: with increasing $\vartheta_\mathrm{M}$ the hydrogen H1 points
more towards the hexagonal axis $\mathbf{c}$ (Fig.~\ref{fig:geom}), which
provides more space for the oscillating part of the molecule---increasing
$\vartheta_\mathrm{M}$ with temperature is thus a plausible consequence of
larger oscillations at high temperatures.    

The best match with the experiment is reached for $\vartheta_\mathrm{M}$ linearly
changing in the range 18.1--19.2$^\circ$ (from low to high temperatures), with
$\frac{\omega_\mathrm{H}}{2\pi}\!\approx\,$\qty{4.95e12}{\hertz} for \w, which
corresponds to a frequency of librations of \qty{165}{cm^{-1}}. As
for the temperature dependence of the
quadrupole splitting in \hw, the best match is reached using
$\frac{\omega_\mathrm{H}}{2\pi}\!\approx\,$\qty{4.34e12}{\hertz}, corresponding to \qty{145}{cm^{-1}}. Both these wave numbers are in a reasonable agreement with values of libration modes of type-I water as observed in optical experiments
\cite{Wood67,Belyanchikov2017}. Out
of the remaining parameters, a part of them is determined by the water molecule geometry: $\delta\!=\,$\qty{58.7}{\kHz}, $d_\mathrm{OH}\!=\,$\qty{0.949}{\angstrom}, and
$\vartheta_{zz}=35.2^\circ$, and the last two parameters do not significantly differ from expectations: \cq\!=\,\qty{176}{\kHz} and $\eta\!=\!0.13$.

We note that the ratio of harmonic frequencies
$\frac{\omega_\mathrm{H}}{\omega_\mathrm{D}}\!\approx\!1.14$ does not reach the
value $\sqrt{2}$, expected from the ratio of deuteron and proton masses. This is
most probably caused by the limitation of our simple oscillator model, where the
amplitudes of oscillations are relatively large at high temperatures. Only a few
lowest energy levels of the oscillator ensemble are populated even at \qty{365}{\kelvin}, as
indicated by the sum of the first five Boltzmann probabilities $\sum_{n=0}^{4}
p_n \approx 0.96$, therefore the difference between quantum harmonic oscillator
model and, e.g., particle in box model would not be large. However, the effect
of walls of the beryl voids should be addressed, especially at high temperatures, by adding some particle-in-box features, which would lead to a more appropriate model such as oscillator confined in a box\cite{Gueorguiev2006}. 

Another limitation of the applied model is that the frequency of linear harmonic oscillator is constant and only the amplitude of motion decreases with temperature. In reality however, the frequency of molecular motions and jumps also reduces on cooling, and at some point the motional averaging becomes inefficient. This is noticeable at temperatures below \qty{70}{\kelvin} where the character of NMR spectra changes and some of the well-resolved doublets become broad bands with features resembling spectra of a powder sample---instead of averaged values the spectra display the actual distributions of the local fields. The narrow peaks with resonance frequencies corresponding to the time-averaged fields are still apparent at \qty{70}{\kelvin}, whereas the NMR spectra at \qty{30}{\kelvin} already contain broad bands (Fig.~\ref{fig:lowT}). In order to understand these spectral shapes we compare them with NMR spectra where the motional averaging was fully or partially removed. Using the spectroscopic and angular parameters obtained from the linear harmonic oscillator model, we constructed the \deu spectrum for $\mathbf{B}_0 \perp \mathbf{c}$ (Fig.~\ref{fig:lowT}a) as a histogram where the angle $\alpha$ has a Gaussian distribution (with dispersion $\alpha_0^2$) and the angle $\varphi_\mathrm{M}$ is uniformly distributed within $\left<(0,2\pi\right)$, i.e., a fully static spectrum. Additionally we simulated a spectrum where the jumps/re-bonding ceased, but the averaging due to librational motion is still effective. Apparently, the experimental spectrum at \qty{30}{\kelvin} contains features of both types of simulated spectra. Similar behavior can be seen for $\mathbf{B}_0 \parallel \mathbf{c}$ (Fig.~\ref{fig:lowT}b), where the outer doublet is broadened as in a fully static spectrum, yet it contains some residual peaks at about 100 and \qty{-100}{\kHz} which stem from the librational averaging. On the other hand, the inner doublet for $\mathbf{B}_0 \parallel \mathbf{c}$ remains unchanged by these effects, which is expected, since it originates from the \deu nuclei of the H2 atoms which have a delta-function-like or very narrow distribution of motions during the librations---as imposed by the directionality of the hydrogen bond. The powder-like features in the low temperature spectra are thus in line with the proposed behavior of water molecules.

\subsection{Analysis of Jumps and Re-bonding of Water Molecules}

The details of the second type of motion of water molecule in the
voids---the re-bonding of hydrogen bond leading to effective
rotations of the water molecules about the beryl hexagonal axis---are more difficult to obtain from our NMR experiments. The twelve bonding sites in the beryl void are equivalent by symmetry and thus cannot be distinguished directly in the NMR spectrum. From the dispersion of
the quadrupole splitting with angle $\varphi_\mathrm{M}$ in case of  $\mathbf{B}_0 \perp \mathbf{c}$ (Fig.~\ref{fig:sensitivity}b) it is apparent that the re-bonding (considered as an instantaneous change in $\varphi_\mathrm{M}$) leads to variations of the splitting $\left<\Delta_\mathrm{Q}\right>_{c}$ for the inner doublet that are about two orders of magnitude larger than those of the outer doublet. Therefore, the outer doublet is well resolved in the \deu NMR spectrum, i.e., the librations and re-bonding lead to sufficient averaging of local fields for the oscillating part of the water molecule (H1 site), whereas the inner doublet is severely broadened when  $\mathbf{B}_0 \perp \mathbf{c}$, which means that the averaging by re-bonding is ineffective. From this we may estimate that the jumps within the beryl void are about two orders of magnitude slower than the molecular librations.

\section{Conclusions}
The experimentally obtained \hyd and \deu NMR spectra, measured at
temperatures 5--\qty{360}{\kelvin}, provided spatial and dynamic information on the behavior
of water molecules in beryl crystal. We showed that although the water
molecules in the undoped beryl, traditionally denoted as type-I molecules,
are indeed oriented preferentially with their H--H lines along the hexagonal axis of beryl, the actual orientation of the H--H lines is significantly (by
about 18$^\circ$) declined from the hexagonal
axis. The reason consists apparently in the fact that the water
molecules are oriented by their O--H bond towards one of twelve
oxygen atoms forming the walls of the beryl structural voids. These
positions are obviously energetically favorable, as hydrogen bonds between the 
\w molecules and the void walls can be formed. In order to explain
satisfactorily our experimental results, we
proposed a model within which the molecules perform two types
of movements: (i) librations
around the axes of such hydrogen bonds, and (ii) less frequent
orientational jumps among the twelve
possible binding sites in the void. A simple thermodynamical model, with
the water molecule considered as a quantum harmonic oscillator,
correctly describes the observed experimental data at all temperatures, for both normal and heavy water. The frequencies of the oscillatory motions, evaluated from our thermodynamic model, agree well quantitatively with the frequencies of libration motions observed by optical experiments\cite{Wood67,Belyanchikov2017}.

The implications of our work can be summarized as follows. First of all, the
traditional view of type-I water molecules embedded in the beryl voids has to
be corrected; a deviation of their H--H lines from the hexagonal axis with a
mean value of about 18$^\circ$ must be accounted for. Consequently, in the
search for ferroelectric order in the water subsystem, it appears important to
bear in mind that, in principle, two components of spontaneous polarization
may exist at low temperatures; a net ferroelectric moment may appear not only
in the $ab$ crystallographic planes but also along the $c$ hexagonal axis.
Concerning the interactions between the water molecules and the atoms forming
the structural voids in beryl, we have provided a very strong indirect
indication that hydrogen bonds are formed between the molecules and any of the
twelve closest oxygen atoms. This contradicts the earlier conjecture by Paré
and Ducros \cite{Pare1964} that no hydrogen bonds between the molecules and
the beryl structure forms. Conversely, the presence of these hydrogen bonds
supports the recent idea of a coupling mechanism between the water
molecules and the beryl crystal lattice, suggested recently by Finkelstein
\textit{et al.}\cite{Finkelstein17}, which agrees very well also with the
observation of sum and difference frequencies in the infrared spectra
\cite{Wood67}. Consequently, our findings, as we believe, can serve as a firm basis for further investigations, including theoretical modeling, of the dynamics of isolated water molecules in beryl, and they can thus also deepen the understanding of mechanisms and interactions favoring the ferroelectric ordering of the water molecules.

\begin{acknowledgments}
This work was supported by the Czech Science Foundation (project No. 20-1527S).    
\end{acknowledgments}

\section*{References}

\bibliography{berylNMR}

\end{document}